\documentclass[a4paper,12pt]{article}
\usepackage{graphicx}
\begin{document}
\bibliographystyle{unsrt}

\newcommand{\beqn}{\begin{eqnarray}}
\newcommand{\eeqn}{\end{eqnarray}}
\newcommand{\ra}{\rightarrow}

\newcommand{\np}{Nucl.\,Phys.\,}
\newcommand{\pl}{Phys.\,Lett.\,}
\newcommand{\pr}{Phys.\,Rev.\,}
\newcommand{\prl}{Phys.\,Rev.\,Lett.\,}
\newcommand{\prep}{Phys.\,Rep.\,}
\newcommand{\nuclinst}{{\em Nucl.\ Instrum.\ Meth.\ }}
\newcommand{\annp}{{\em Ann.\ Phys.\ }}
\newcommand{\intjmp}{{\em Int.\ J.\ of Mod.\  Phys.\ }}


\newcommand{\mw}{M_{W}}
\newcommand{\mww}{M_{W}^{2}}
\newcommand{\mwmw}{M_{W}^{2}}

\newcommand{\mz}{M_{Z}}
\newcommand{\mzz}{M_{Z}^{2}}

\newcommand{\cw}{\cos\theta_W}
\newcommand{\sw}{\sin\theta_W}
\newcommand{\tw}{\tan\theta_W}
\def\cww{\cos^2\theta_W}
\def\sww{\sin^2\theta_W}
\def\tww{\tan^2\theta_W}

\def\noi{\noindent}
\def\nn{\noindent}

\def\sinb{\sin\beta}
\def\cosb{\cos\beta}
\def\sinbb{\sin (2\beta)}
\def\cosbb{\cos (2 \beta)}
\def\tgb{\tan \beta}
\def\tgbt{$\tan \beta\;\;$}
\def\tgbsq{\tan^2 \beta}
\def\sel{\tilde{e}_L}
\def\ser{\tilde{e}_R}
\def\msel{m_{\sel}}
\def\mser{m_{\ser}}
\def\mslr{m_{\tilde{l}_R}}
\def\m0{M_0}
\def\amu{\delta a_\mu}

\def\slashE{E\kern -.620em {/}}

\def\mchi{m_\chi^+}
\def\neuto{\tilde{\chi}_1^0}
\def\mneuto{m_{\tilde{\chi}_1^0}}
\def\neutt{\tilde{\chi}_2^0}
\def\neutth{\tilde{\chi}_3^0}
\def\ma{M_A}
\def\mstau{m_{\tilde\tau}}
\def\msne{m_{\tilde\nu}}
\def\msnee{m_{{\tilde\nu}_e}}
\def\mh{M_h}

\def\sinb{\sin\beta}
\def\cosb{\cos\beta}
\def\sinbb{\sin (2\beta)}
\def\cosbb{\cos (2 \beta)}
\def\tgb{\tan \beta}
\def\tgbt{$\tan \beta\;\;$}
\def\tgbsq{\tan^2 \beta}
\def\sinal{\sin\alpha}
\def\cosal{\cos\alpha}
\def\stop{\tilde{t}}
\def\sto{\tilde{t}_1}
\def\stt{\tilde{t}_2}
\def\stl{\tilde{t}_L}
\def\str{\tilde{t}_R}
\def\msto{m_{\sto}}
\def\mstosq{m_{\sto}^2}
\def\mstt{m_{\stt}}
\def\msttsq{m_{\stt}^2}
\def\mt{m_t}
\def\mtsq{m_t^2}
\def\sint{\sin\theta_{\stop}}
\def\sintt{\sin 2\theta_{\stop}}
\def\cost{\cos\theta_{\stop}}
\def\sintsq{\sin^2\theta_{\stop}}
\def\costsq{\cos^2\theta_{\stop}}
\def\mqtt{\M_{\tilde{Q}_3}^2}
\def\mutt{\M_{\tilde{U}_{3R}}^2}
\def\sbottom{\tilde{b}}
\def\sbo{\tilde{b}_1}
\def\sbt{\tilde{b}_2}
\def\sbl{\tilde{b}_L}
\def\sbr{\tilde{b}_R}
\def\msbo{m_{\sbo}}
\def\msbosq{m_{\sbo}^2}
\def\msbt{m_{\sbt}}
\def\msbtsq{m_{\sbt}^2}
\def\mt{m_t}
\def\mtsq{m_t^2}
\def\selectron{\tilde{e}}
\def\seo{\tilde{e}_1}
\def\set{\tilde{e}_2}
\def\sel{\tilde{e}_L}
\def\se1{\tilde{e}_1}
\def\ser{\tilde{e}_R}
\def\mseo{m_{\seo}}
\def\mseosq{m_{\seo}^2}
\def\mset{m_{\set}}
\def\msetsq{m_{\set}^2}
\def\msel{m_{\sel}}
\def\mser{m_{\ser}}
\def\mse1{m_{\se1}}
\def\me{m_e}
\def\mesq{m_e^2}
\def\snu{\tilde{\nu}}
\def\snue{\tilde{\nu_e}}
\def\set{\tilde{e}_2}
\def\snul{\tilde{\nu}_L}
\def\msnue{m_{\snue}}
\def\msnuesq{m_{\snue}^2}
\def\smuon{\tilde{\mu}}
\def\smul{\tilde{\mu}_L}
\def\smur{\tilde{\mu}_R}
\def\msmul{m_{\smul}}
\def\msmulsq{m_{\smul}^2}
\def\msmur{m_{\smur}}
\def\msmursq{m_{\smur}^2}
\def\stau{\tilde{\tau}}
\def\stauo{\tilde{\tau}_1}
\def\staut{\tilde{\tau}_2}
\def\staul{\tilde{\tau}_L}
\def\staur{\tilde{\tau}_R}
\def\mstauo{m_{\stauo}}
\def\mstauosq{m_{\stauo}^2}
\def\mstaut{m_{\staut}}
\def\mstautsq{m_{\staut}^2}
\def\mtau{m_\tau}
\def\mtausq{m_\tau^2}
\def\gluino{\tilde{g}}
\def\mgluino{m_{\tilde{g}}}
\def\mchi{m_\chi^+}
\def\neuto{\tilde{\chi}_1^0}
\def\mneuto{m_{\tilde{\chi}_1^0}}
\def\neutt{\tilde{\chi}_2^0}
\def\mneutt{m_{\tilde{\chi}_2^0}}
\def\neutth{\tilde{\chi}_3^0}
\def\mneutth{m_{\tilde{\chi}_3^0}}
\def\neutf{\tilde{\chi}_4^0}
\def\mneutf{m_{\tilde{\chi}_4^0}}
\def\chargop{\tilde{\chi}_1^+}
\def\chargopm{\tilde{\chi}_1^\pm}
\def\mchargo{m_{\tilde{\chi}_1^+}}
\def\chargtp{\tilde{\chi}_2^+}
\def\mchargt{m_{\tilde{\chi}_2^+}}
\def\chargom{\tilde{\chi}_1^-}
\def\chargtm{\tilde{\chi}_2^-}
\def\bino{\tilde{b}}
\def\wino{\tilde{w}}
\def\photino{\tilde{\gamma}}
\def\zino{tilde{z}}
\def\sdowno{\tilde{d}_1}
\def\sdownt{\tilde{d}_2}
\def\sdownl{\tilde{d}_L}
\def\sdownr{\tilde{d}_R}
\def\supo{\tilde{u}_1}
\def\supt{\tilde{u}_2}
\def\supl{\tilde{u}_L}
\def\supr{\tilde{u}_R}
\def\mh{m_h}
\def\mht{m_h^2}
\def\MH{M_H}
\def\MHt{M_H^2}
\def\MA{M_A}
\def\MAt{M_A^2}
\def\MHp{M_H^+}
\def\MHm{M_H^-}
\def\epem{e^+e^-}
\def\epemt{$e^+e^-$}
\def\siginv{\sigma_{\gamma+inv}}
\def\gmuon{$(g-2)_\mu$}
\def\r12{r_{12}}
\def\bsgamma{b\ra s\gamma}
\def\bsmu{B_s\ra \mu^+\mu^-}
\def\feynhiggs{{\tt FeynHiggsfast}}

\begin{titlepage}
\def\baselinestretch{1.2}
\vspace*{\fill}
\begin{center}
{\large {\bf {\em Lower limit on the neutralino mass in the
general MSSM.}}} \vspace{1cm}

\begin{tabular}[t]{c}

{\bf G.~B\'elanger$^{1}$, F.~Boudjema$^{1}$, A.~Cottrant$^{1}$, A. Pukhov$^{2}$, S. Rosier-Lees$^{3}$}

\\
\\

{\it 1. Laboratoire de Physique Th\'eorique}
{\large
LAPTH},\\
 {\it Chemin de Bellevue, B.P. 110, F-74941 Annecy-le-Vieux,
Cedex, France.}\\
2. {\it Skobeltsyn Institute of Nuclear Physics, Moscow State University},\\
{\it Moscow, Russia.}\\
3.{\it  Laboratoire d'Annecy-le-Vieux de Physique des Particules}
{\large LAPP},\\
 {\it Chemin de Bellevue, B.P. 110, F-74941 Annecy-le-Vieux,
Cedex, France.}\\
 \\

\end{tabular}
\end{center}

\centerline{ {\bf Abstract} } \baselineskip=14pt \noindent
{\small We discuss constraints on SUSY models with non-unified
gaugino masses and $R_P$ conservation. We derive
 a lower bound on the neutralino mass
combining  the direct limits  from LEP, the indirect limits from
\gmuon, $\bsgamma$, $B_s\ra \mu^+ \mu^-$  and the relic density
constraint from {\tt WMAP}. The lightest neutralino
$\mneuto\approx 6$GeV is found in models with a light pseudoscalar
with $\ma<200$GeV and a large value for $\tan\beta$. Models with
heavy pseudoscalars lead to $\mneuto>18(29)$GeV for
$\tan\beta=50(10)$. We show that even a very conservative bound
from the muon anomalous magnetic moment can  increase the lower
bound on the neutralino mass in models with  $\mu<0$ and/or large
values of $\tan\beta$. We then examine the potential of the {\tt
Tevatron}  and the direct
 detection experiments to probe the SUSY models with the lightest
neutralinos allowed in the context of light pseudoscalars with
high $\tgb$. We also examine the potential of an $e^+e^-$ collider
of 500GeV to produce SUSY particles in all models with neutralinos
lighter than the $W$. In contrast to the mSUGRA models,
observation of at least one sparticle is not always guaranteed.}
\vspace*{\fill}

\vspace*{0.1cm} \rightline{LAPTH-1001/03}

\rightline{{\today}}

\end{titlepage}
\baselineskip=18pt
\baselineskip=14pt



\section{Introduction}

The minimal supersymmetric standard model (MSSM) is certainly one
of the most attractive extensions of the standard model. Apart
from providing a solution to the hierarchy problem it also fits in
very well with the idea of the gauge unification and gives as a
bonus a good dark matter candidate represented by its lightest
supersymmetric particle, the LSP for short. One the other hand, to
the question of how light this  LSP can be so that one has an idea
about the masses for the rest of the spectrum, the model provides
no answer. The reason for this is that there is no definite model
for the masses of the sparticles (and other scales) which would
ultimately point out to a theory of the (soft) breaking of
supersymmetry. One exception relates to the mass of the Higgs if
one makes the mild requirement that the scale of supersymmetry
breaking to be much above the electroweak scale. This gives one of
the robust predictions of the model which is that there must be at
least one light Higgs with mass below about $130$GeV
\cite{mssmmhmax}, for alternatives see
\cite{hardterms-higgs-review,me-3h-2hdm}. For practically all the
other (s)particles of the model one does now have some lower
bounds on their masses from extensive phenomenological studies and
searches both at colliders and in astroparticle experiments. One
can for example mention the impact of both {\tt LEP} and the
{\tt Tevatron} in constraining the model parameters through the direct
searches, the importance of rare decays notably $b\ra s \gamma$
or even low energy experiments like the recent measurement of the
muon anomalous moment \gmuon~\cite{g-2ex}. Cosmological
considerations like the thermal relic density of the LSP also put
some stringent bounds on the MSSM. It should however be emphasised
that many of the bounds and constraints, especially those from
indirect searches, but not only, are based on theoretical
assumptions some of which with strong prejudices. The aim of this
paper is to critically review the lowest bound on the neutralino
LSP with as little  theoretical bias as possible. Instead of
reviewing a host of models, we will try to delimit the parameter
space that has an incidence on the LSP mass bound in a, as much as
possible, model independent way \cite{nous-susy02}. We will see
that one does not have to play with a large number of parameters
to arrive at the lowest bound on the LSP, instead the parameter
space will be naturally reduced to only a few critical parameters.

A large amount of work has been devoted to the so-called minimal
supergravity model (mSUGRA) mainly because of its simplicity.
Indeed, instead of dealing with about $100$ parameters in the
general MSSM, one works with only four parameters and a sign. With
such a constrained model it becomes possible to combine the
various experimental data and severely delimit the parameter
space. Within this model the lowest limit on the LSP is extracted
in a straightforward way from the {\tt LEP} data which give
$\mneuto>59$GeV \cite{lep_neutralino,lep_susy_WG}. Moreover, with
the new precise determination of the relic density of cold dark
matter by {\tt WMAP} \cite{wmap}, the mSUGRA parameter space
considerably shrinks to thin strips \cite{wmap-sugra} suggesting a
rather focused phenomenology and benchmarks for the next colliders
\cite{benchmarks-postwmap}. Supersymmetric signals can be
radically different in other scenarios, scenarios where one can
have much more room for manoeuvre even after imposing all the
latest data. In particular, the LSP bound can be much smaller than
the one found in the context of mSUGRA as we will see.
\\
\noi Different approaches have been followed aiming at a reduction
of parameters. Theoretically motivated approaches include the
so-called anomaly mediated (AMSB) models \cite{AMSB} and
generalisations \cite{non-universal-anomaly} for example, or
schemes that appeal to superstrings that are either
moduli-dominated or contain a mixture of moduli and dilaton fields
\cite{nonuniversal-strings,multimoduli,CDG,Ben-Quevedo} with
varying degrees of simplification. There are even scenarios that
interpolate between mSUGRA and some of the AMSB and string
inspired methodology \cite{binet-nonuni}. Another approach without
necessarily an underlying  framework for  supersymmetry breaking,
lifts, at the GUT scale, the mass degeneracies of the mSUGRA model
either in the sector of the gaugino
\cite{nmSUGRA,non-universal-su5,baernonuni,orloff-nonuni,nonuniversal-birkedal},
Higgs \cite{higgsmass-nonuni,arnowittnonuni} or the sfermions
\cite{gomez-nonuni,orloff-nonuni}. Some analyses let loose the
shackles of mSUGRA and models defined at high scale but still
impose ad-hoc constraints, like equality of scalar masses at the
electroweak scale \cite{uni-ewscale}. Reviewing this situation one
can still ask about the lowest {\em stable} LSP mass allowed by
the present data regardless of a pre-defined scheme. This is
important because not only one would want to know what the impact
of the lightest LSP is for cosmology and direct detection but how
the collider phenomenology can get affected when one puts aside
assumptions related to the LSP mass. For example one might inquire
about the sensitivities of the direct search detectors such as
{\tt CDMS} \cite{cdms}, {\tt EDELWEISS} \cite{edelweiss} and {\tt
ZEPLIN} \cite{zeplin} for neutralino masses below those typical of
the LSP of mSUGRA. The LSP, in the R-parity conserving MSSM, being
the end product of the decay of {\em any} supersymmetric particle,
is important for signatures at the colliders since it will always
show up in any analysis.
 Another reason why it is interesting to study the light neutralino is that
it   opens the door for a sizable branching fraction of the Higgs
into invisible thus reducing all the branching fractions into the
usual discovery channels, $h\ra \gamma\gamma$ or $h\ra b\bar{b}$
\cite{nous_hinvisible_lhc,oursusyg-2}. This possibility has
triggered analyses by the LHC experiments to search for the Higgs
with a sizable branching fraction into invisible
\cite{hinvisble-lhc-ex,Zeppenfeld-h-invisible}. This has also
motivated studies to reanalyse {\tt Tevatron}  data in models with
light neutralinos and establish whether or not supersymmetry could
be discovered there in the chargino/neutralino channel leading to
tri-leptons \cite{tri-leptons-nonuni}.

The oft-quoted lower limit, $\mneuto>59$GeV
\cite{lep_neutralino,lep_susy_WG} which applies to mSUGRA, is
basically derived from the quite robust model independent lower
limit on the chargino mass $\mchargo$. The (lightest) chargino
mass  is set from chargino pair production at LEP2. Deriving the
lower limit on the neutralino mass from this analysis, on the
other hand, tacitly assumes a model of unified gaugino masses at
the GUT scale, much like what is hypothesised in mSUGRA. However
in a general MSSM, the masses $\mneuto$ and $\mchargo$ are
uncorrelated and the lower limit on the LSP neutralino mass
weakens considerably if only the {\tt LEP} constraint is taken
into account. This occurs for a bino neutralino when $M_1 \ll M_2$
where $M_1$($M_2$) is the $U(1)$($SU(2)$) soft-susy breaking
gaugino mass. To derive a lower bound on the LSP in these
conditions one has to turn to cosmology. Indeed, the neutralino
LSP cannot be too light as it would conflict  with the precise
measurement of the thermal relic density of cold dark matter
inferred from {\tt WMAP} \cite{wmap}. One is then indirectly led
to include the sleptons into the picture, in particular the right
handed ones, $\tilde{l}_R$. Indeed a light neutralino that is
mostly a bino will annihilate preferably into fermions through
right-handed sleptons because the latter  have the largest
hypercharge among all sfermions
\cite{relic-classic,astrosusy-review}. As a rule of thumb, with
all sfermions heavy but the three right sleptons, a rough
approximate requirement is
\beqn
\label{approx-relic} m_{\tilde{l}_R}^2 < 10^{3} \sqrt{(\Omega_\chi
h^2)_{\rm max}} \times \mneuto.
\eeqn
with all masses expressed in GeV. We identify $\Omega_\chi$ with
the fraction of the critical energy density provided by neutralino
LSP and $h$ is the Hubble constant in units of 100 km sec$^{-1}$
Mpc$^{-1}$. Therefore the lower bound on the slepton from LEP2,
when  used in conjunction with the upper limit from the relic
density\footnote{Note that we  only consider  scenarios where the
neutralino contributes to the cold dark matter. Unstable
neutralinos with masses in the MeV range have been entertained
\cite{dreiner_karmen} and their astrophysical (supernova)
implications studied.}, plays an important role in constraining a
light neutralino. We will see that to accommodate the relic
density constraint, a slight Higgsino component can make the
mostly bino LSP couple to the $Z$ and the Higgses. If one is not
far from these resonances there will be efficient annihilation of
the neutralino LSP.

To further constrain the lower bound on the LSP mass and bring in
new observables in the picture, some assumptions on the sfermion
masses need to be made.  A well motivated mild assumption relies
on so-called minimal flavour violation in order to evade bounds on
flavour changing neutral currents. In the context of a generic
minimal flavour violation model, sfermions with the same quantum
numbers would share the same mass, with perhaps a slight breaking
from Yukawa mixing. We would then treat all the sleptons on an
equal footing. This then brings into the picture the constraint
from the measurement of the muon anomalous magnetic moment,
$\delta a_\mu$. The latter
 is also sensitive to the presence of light charginos/neutralinos and smuons.
Both the theoretical predictions and experimental results on the
muon anomalous magnetic moment have been refined on several
occasions in the last year and the situation is still evolving. We
will therefore be very conservative in imposing limits from \gmuon~
using numbers that encompass different estimates. Nevertheless we
will find that the upper bound on the \gmuon~ does constrain some
of our scenarios. In our study we will also show how results
change if one does not impose the \gmuon~ bound.

The squarks on the other hand hardly enter the game. One needs
however to assume them to be sufficiently heavy so that,
especially for large $\tan\beta$, one does not conflict with the
bound from $b\ra s \gamma$. Moreover, squarks of the third
generation should be heavy enough and or the tri-linear mixing
parameter of the top be large so that,   especially for low
$\tan\beta$, one evades more easily the lower bound on the Higgs
mass. One usually also chooses the pseudoscalar Higgs mass to be
heavy enough to achieve this. With the assumption that the
pseudoscalar Higgs is heavy (beyond say $\sim 300$GeV), an
absolute lower limit on the neutralino mass,
$\mneuto=18$GeV($29$GeV) for $\tan\beta=50\;(10)$ results from
combining cosmological and collider constraints.

Relaxing the assumption that the  pseudoscalar is very heavy,
after all the {\tt LEP} lower bound is only about $90$GeV, leads
to an even smaller lower bound on the neutralino mass,
$\mneuto>6$GeV. This is because, as was pointed out in
\cite{Torino-lsp2}, a new possibility  for escaping the relic
density constraint for very light neutralinos opens up: neutralino
annihilation into $b\bar{b}$ pairs via a Higgs resonance. This
channel becomes efficient in the large $\tan\beta$ regime due to
the enhanced couplings to $b$ quarks and $\tau$'s. We will show
that the concomitant presence of a light charged Higgs at large
$\tan\beta$ means that these models are tightly constrained by
$\bsgamma$ as well as by $\bsmu$. Nevertheless there remains a
possibility for very light neutralinos, $\mneuto\sim 6$GeV, when
$\tan\beta>25$. The phenomenological implications of these classes
of models is markedly different from the models with heavy
pseudoscalars and deserves special attention. Several Higgses with
masses in the
 $100-200$GeV range and with high \tgbt
can be detected at the {\tt Tevatron} , even with $2 fb^{-1}$.
These models are testable at the current RunIIa. In this scenario,
the Higgs exchange also dominates the spin-independent
neutralino-nucleon cross-section. Large cross-sections are found
for very light neutralinos, $\mneuto\sim 6-10$GeV.

This paper is organised as follows. We first critically review the
constraint from LEP, paying attention both to the limits on
neutralinos, charginos as well as sleptons. We will then analyse
the constraints from the relic density and from the muon anomalous
magnetic moment, $(g-2)_\mu$ and the rare decay $b\to s\gamma$.
Other constraints that we use, although in most cases without much
impact, are the effect on the invisible width of the $Z$, $Z\ra b
\bar b$ and the rare decay $B_s\ra \mu^+ \mu^-$.  The latter can
play a r\^ole for a light pseudoscalar. We will concentrate on the
intermediate to large $\tan\beta$ region and present absolute
lower limits on the mass of the neutralino LSP. The case of models
with a light pseudoscalar Higgs deserves a special section both as
concerns the impact of the various constraints as well as the
prospects at future colliders. Here the {\tt Tevatron}  plays an
important role. The prospects for direct detection experiments
will be discussed in this  context. Finally, a section will be
devoted to the potential of the linear collider to produce
supersymmetric particles in models with light neutralinos,
$\mchargo<M_W$. We will close with a  conclusion that  summarises
our results.

\section{MSSM parameters}

\subsection{Physical parameters}
When discussing the physics of charginos and neutralinos it is
best to start by defining one's notations and conventions. All our
parameters, unless stated otherwise,  are defined at the
electroweak scale. The chargino mass matrix in the
gaugino-Higgsino basis is defined as
\beqn
\label{charginomatrix} \left(
\begin{array}{cc}
M_2 & \sqrt{2} \mw \cosb  \\ \sqrt{2} \mw \sinb   & \mu
\end{array}  \right)
\eeqn
\noindent where $M_2$ is the soft SUSY breaking mass term for the
$SU(2)$ gaugino while $\mu$ is the so-called Higgsino mass
parameter whereas $\tgb$ is the ratio of the vacuum expectation
values for the up and down Higgs fields.

Likewise the neutralino mass matrix is defined as
\beqn
\label{neutralinomatrix} \left(
\begin{array}{cccc}
M_1 & 0 & -\mz \sw \cosb & \mz \sw \sinb \\ 0 & M_2 & \mz \cw
\cosb & -\mz \cw \sinb \\ -\mz \sw \cosb   & \mz \cw \cosb  & 0 &
-\mu \\ \mz \sw \sinb & -\mz \cw \sinb & -\mu & 0
\end{array}  \right)
\eeqn
\noi where the first entry $M_1$ (corresponding to the bino
component) is the $U(1)$ gaugino mass. The oft-used gaugino mass
unification condition corresponds to the  assumption
\beqn
\label{m1m1unification} M_1=\frac{5}{3} \tan^2\theta_W M_2 \simeq
M_2/2
\eeqn
Then constraints from the charginos alone can be easily translated
into constraints on the neutralino sector. Relaxing
Eq.~\ref{m1m1unification}, or removing any relation between $M_1$
and $M_2$ means that one needs further observables specific to the
neutralino sector. It is a trivial observation that if $M_1\ll
M_2, |\mu|$ one can get a very low neutralino mass independently
of the chargino mass derived from Eq.~\ref{charginomatrix}.

In the approach we are taking, the free parameters include the ones of the
 gaugino sector as well as the parameters of the slepton sector.
As discussed earlier, we will assume that all the squarks are
heavy. Allowing light squarks would not affect much the
cosmological constraints for a neutralino that is mostly bino
 and in any case light squarks are strongly constrained from $\bsgamma$.
 For the slepton sector we will consider primarily  a model reminiscent of
 mSUGRA models but without gaugino mass unification.
This model  features  a common
 mass for the sleptons at the GUT scale. As we will see, for
 the purpose of constraining the LSP mass, the slepton mass spectrum is
 essentially the same had we imposed the more general requirement
 of minimal flavour violation by giving a common mass to the left
 and a common mass to right sleptons of all three generations.

Using the renormalisation group equations, the masses at the weak
scale can be related to the ones at the GUT scale. For sleptons
this can be done rather independently of the other MSSM
parameters. Defining the parameter
\beqn
\r12=\frac{M_1}{M_2},
\eeqn
which characterises the amount of non universality in the gaugino
masses, the weak scale slepton masses write
\beqn
\label{m0running} \mser^2&=&\m0^2\;+\; .88 \;\r12^2 M_2^2\;-\;\sww
D_z \nonumber
\\ \msel^2&=&\m0^2\;+ (0.72+.22\; \r12^2) M_2^2\; -\;(.5-\sww)D_z \nonumber
\\ m_{{\tilde \nu}_e}^2&=&\m0^2\;+ (0.72+.22 \;\r12^2) M_2^2\;+\;D_z/2
\;\;\;\;\;\; {\rm with} \nonumber
\\  D_z&=&\mzz \cosbb
\eeqn
Here the gaugino masses, $M_{1,2}$ are defined at the electroweak
scale. To obtain $M_1=\r12 M_2$ at the electroweak scale one needs
$\overline{M_1}\approx 2 \r12 \overline{M_2}$ where
$\overline{M_i}$ are defined at the GUT scale and $M_2\approx
0.825\overline{M_2}$. In the mSUGRA models one has
$\overline{M_1}=\overline{M_2}$. The parameter $\m0$ is here
defined at the GUT scale. For the sake of simplicity we have
neglected the Yukawa couplings in the RGE that could affect the
$\tilde{\tau}$ sector, however  mixing will be taken into
account through the $\mu$ term as we will see.\\
\noi We will be mainly interested in models where $\r12<<1$, then
in the RGE equation all terms in $\r12$ will be negligible. One
obtains a natural splitting between the right-handed/left-handed
sfermion masses.  Indeed, $\mser\approx\m0$ and  is typically much
lighter than $\sel$ which receives in addition a contribution,
$\msel^2\propto M_2^2$.

 The
non universality in the GUT-scale relation for gauginos which we
investigate in this paper are quite plausible as many models
beyond mSUGRA feature some kind of non-universal masses. For
example SUGRA models with non-minimal kinetic terms
\cite{nmSUGRA}, superstring models with moduli-dominated or a
mixture of moduli and dilaton fields
\cite{nonuniversal-strings,multimoduli,non-universal-anomaly,CDG,binet-nonuni}
and anomaly mediated SUSY breaking models \cite{AMSB} all feature
non-universal masses in the gaugino and/or scalar masses.

\noi Note in passing that Eq.~\ref{m0running} can be extended to
squarks and if we take $\overline{M}_3=r_{32} \overline{M}_2$ with
$r_{32}>1$ at the GUT scale one could make the squarks ``naturally
heavy" as we have assumed. Then the gluino mass relation
$m_{\tilde{g}}\sim 4 M_2$ obtained in mSUGRA type models turns
into $m_{\tilde{g}}\sim 4 r_{32} M_2$ ($r_{32}>1$). First and
second generation squark masses, neglecting the small bino
contribution $\propto M_1^2$, can be approximated as
\beqn
\label{msquarkrunning} m_{\tilde{q}_{L,R}}^2 \sim
m_{\tilde{e}_{L,R}}^2 + 0.6 m_{\tilde{g}}^2
\eeqn

We fill first consider the case where the pseudoscalar is heavy, $\ma=1$~TeV.
Altogether we allow 6 free parameters at the electroweak scale and fix
 $M_3=m_{\tilde q}=1$~TeV :
 \beqn
 \tan\beta,M_1,M_2,\m0,\mu,A_q
 \eeqn
$A_{q,l}$ are the tri-linear terms for the quark $q$ and lepton
$l$.  For scans over the parameter space, unless otherwise
specified, we will consider the range
 \beqn \label{scan}
5<\tan\beta<50,\nonumber\\
 M_2<2000~GeV,\nonumber\\
  .001<\r12<.6,\nonumber\\
  |\mu|<1000~GeV,\nonumber\\
 |A_t|<2400~GeV,\nonumber\\
   \m0<1000~GeV.
  \eeqn

\noi We will  fix $A_l=0$ for all sleptons as most of the
processes we will discuss are not very sensitive to the exact
value of this parameter. Although the mixing in the stau sector,
 which is $\propto \tilde{A_\tau}=A_\tau-\mu\tan\beta$,
 can be relevant for the calculation of the relic density,
 in our case  we always have $|\mu|\tan\beta>1000$
so  this term usually dominates the mixing.
Finally we will consider also the case where the pseudoscalar
mass is a free parameter varying it in the range
\beqn
92 GeV<\ma<1TeV.
\eeqn

\section{Limits from LEP}

The direct limits from LEP2 on charginos, neutralinos as well as
on sleptons are relevant for the lower bound on the   lightest
neutralino. As argued, the sleptons play an important role in the
relic density calculation especially when the light neutralino is
a bino. Here we revisit the lower limits that can be obtained by
LEP2 on sfermions and gauginos/Higgsinos when one relaxes the
gaugino universality assumptions.

\noi
{\bf Charginos}

The lower bound on the chargino mass rests near the kinematic
limit, $\mchi> 103.5$GeV unless the sneutrino lies in the range
$75<\msnee<85$GeV, then the bound drops to $\mchi>73$
GeV \cite{lep_susy_WG}. This is due  to the destructive
interference between the
 $t$ and $s$- channel contributions to $\epem\ra\chargop\chargom$.
 These bounds basically translate into bounds on $(M_2,|\mu|)>(73,103)$GeV.

\noi
{\bf Neutralinos}

The {\tt LEP} experiments quote a  lower limit on the neutralino
mass, $\mneuto <59.6GeV$, while  assuming  unified  gaugino masses
at the GUT scale, $\r12=0.5$ \cite{lep_neutralino}. This
constraint on the neutralino mass is basically derived from the
lower limit on the chargino mass obtained in the pair production
process, $\epem\ra\chargop\chargom$ which depends on $M_2,\mu$. It
is only through the relation between the gaugino parameters $M_1$
and $M_2$ that the neutralino mass limit is obtained. In a general
MSSM, the charginos and neutralino  masses are uncorrelated and
the lower limit on the neutralino mass weakens when $\r12<.5$
becoming roughly $\mneuto\geq r_{12}\times\mchargo$. Taking into
account the constraint from the chargino sector, the lightest
neutralinos are then typically found in scenarios where $M_1 \ll
M_2,\mu$, that is the light LSP mass is set by the parameter $M_1$
and is thus  mostly a bino.

In such scenarios, the processes $\epem\ra \chi_1^0\chi_2^0,
\chi_1^0\chi_3^0$ can be used to somewhat constrain the parameter
space. In particular, when sleptons are light, the neutralino
cross-sections depends crucially on the Higgsino content. The
cross-sections are generally enhanced and regions with small $\mu$
can be more severely constrained by the neutralino production
cross sections than by the chargino process
\cite{nous_hinvisible_lhc}. In our scans, we have implemented the
upper limit from the {\tt L3} experiment on
$\sigma(\epem\ra\neuto\neutt+\neuto\neutth\ra\slashE  l^+l^-)$, with 
$l=e,\mu$,
using the tables in \cite{L3_susy}
that simulate both the signal and background.

The radiative processes where
a photon is emitted in addition to a pair of invisible
supersymmetric particles, like the lightest neutralino,  will contribute to the
process $\epem\ra \gamma+invisible$ which has been searched for by
the LEP2 experiments.
The neutralino radiative process  $\epem\ra \neuto\neuto\gamma$,  is essentially dominated by the
$t$-channel selectron exchange diagrams and has a small
cross-section \cite{pandita_photino,dreiner_karmen}. We computed
the double differential cross-section(energy and angular
distribution of the photon) as well as the total cross-section
exactly using calcHEP \cite{calchep} for a wide range of parameters
in the MSSM. The  cross-section can reach  $\sigma=50fb$ for
$\mser\approx 100$GeV but drops steadily as the scalar mass
increases. While a few events are expected at LEP2, this is not
enough to overcome the uncertainty in the standard model neutrino
contribution to the $\gamma$+invisible channel. Only mild
improvement is obtained from fitting to the full energy-angular
distribution as the distribution is somewhat similar to the
background, peaked near low-energy photon.
We conclude that one cannot constrain any further the
neutralino mass using this process independently of the selectron mass.

In principle the upper limit on the invisible width of the $Z$,
$\Gamma_{Z_{inv}}<3MeV$ from LEP1 can also constrain the lightest
neutralino.  In the particular case of a bino LSP with $M_1 \ll
M_2$,  the coupling to the $Z$ strongly depends on the amount of
Higgsino mixing.  Thus one obtains a lower bound on $\mu$,  in the
large $\tan\beta$ limit, $|\mu|> 110$GeV.
 This constraint is in general  already satisfied after taking into account
  the LEP2 data on chargino/neutralino production.

\noi
{\bf Sfermions}

For selectrons, a limit of 99.5GeV can be set on both $\sel$ as well as $\ser$
in the case of a light neutralino,  whereas
 basically model independent
limits of $m_{\smuon} > 96$GeV  and  $m_{\stau} > 86$GeV can be
reached \cite{lep_susy_WG}. It is important to point out that the
lower limit on the stau is about $10$GeV smaller compared to the
smuon and the selectron as this will have a consequence on the
relic density contribution. Note also that because of the mixing
is stau sector (which in our case is induced solely through the
$\mu$ term), we can with the same minimum value of $M_0$ in
Eq.~\ref{m0running} arrive at lower values for staus than for
selectrons.

\noi
{\bf Higgs}

In  models where the pseudoscalar mass is heavy,
 the limit on the lightest CP-even Higgs mass from LEP2, $M_h>114.4$ GeV, applies.
We have used \feynhiggs~\cite{FeynHiggs} to calculate the Higgs mass and have
imposed the limit $M_h>113$ GeV to allow for theoretical
uncertainties.

In models with a light pseudoscalar the above LEP2 constraint is
relaxed. When $M_h\approx M_A$ and $\cos(\alpha-\beta)\approx 1$,
the channel $\epem\ra hZ$ is strongly suppressed. LEP2 can only
make use of the $hA\ra bbbb,\tau\tau b b$ channels to put an
absolute bound of $M_h,M_A > 91.6$~GeV
\cite{aleph_HA,delphi_HA,combined_HA}. In these models, the heavy
CP-even Higgs channel $\epem\ra HZ$ which is $\propto
\sin^2(\alpha-\beta)$  is favoured. Unfortunately the mass
splitting between the two scalar Higgses ($M_H>M_A$) is generally
sufficient to put the heavy Higgs beyond the reach of LEP2.

\section{Indirect  limits : $\Omega h^2$, \gmuon, $b\ra s\gamma$,
$B_s \ra \mu^+ \mu^-$ and $Z \ra b \bar b$}

\subsection{Relic density of neutralinos}

The  MSSM model with a light stable neutralino
 must be consistent with at least  the
 upper limit on the amount of cold dark matter.
Here we take the latest bound from {\tt WMAP}, $\Omega h^2< .128$
and also compare with the old limit (of 2001) from {\tt BOOMERANG}
\cite{Boomerang2001}, {\tt MAXIMA} \cite{Maxima2001} and {\tt
DASI} \cite{Dasi2001} with $\Omega h^2< 0.3$. We will refer to this
limit as pre-{\tt WMAP}. Our calculations of the relic density is
based on {\tt micrOMEGAs} \cite{micromegas}, a program that
calculates the relic density in the MSSM including all possible
coannihilation channels \cite{micromegas}. For the light
neutralino masses under consideration,
 it is  the main neutralino annihilation channels that are most relevant, in particular
 annihilation  into a pair of light fermions.
 Basically two types of diagrams contribute, $s$-channel $Z$ (or Higgs) and
$t$-channel sfermion exchange. A light neutralino that is mainly a
bino couples preferentially to right-handed sleptons, the ones
that have the largest hypercharge. To have a large enough
annihilation rate (in order to bring down the relic density below
the upper limit allowed) one needs either a light slepton or a
mass close to $M_{Z,h,H,A}/2$. In the former case, the constraint
from {\tt LEP} on sleptons and in particular staus plays an
important role. In the heavy slepton case,  the coupling of the
$Z$ should be substantial, which requires that the neutralino
should have a certain Higgsino component
\cite{nous_hinvisible_lhc,oursusyg-2}. This means $\mu$ small, but
still consistent with the  chargino constraint.

\subsection{$(g-2)_\mu$}

To derive the bound on $\delta a_\mu=a_\mu^{{\rm
exp.}}-a_\mu^{{\rm theo}}$ we take into account the recent
correction to the sign of the hadronic light by light
contribution, $a_\mu^{LBL}$ \cite{g-2Knecht,g-2correctingerror}. Considering that
there are still a few issues that need to be clarified, see for
instance \cite{g-2Wise}, we allow a larger theoretical error and
take like in \cite{g-2Nyffeler} $a_\mu^{LBL}=(+80\pm 40)\;\times
10^{-11}$. For the hadronic vacuum polarisation we take an
average of the new results. Jegerlehner has reported a new value
that includes BES, CMD-2 with a value on the hadronic part
$a_\mu^{had}=(6889\pm 58)\times
10^{-11}$ \cite{g-2-jegerlehner2003}. Teubner
\cite{g-2-teubner2003} quotes two values (inclusive and exclusive,
depending how the data is integrated) but uses QCD sum rules to
favour the value extracted from inclusive data,
$a_\mu^{had}=6831\pm 59\pm 20)\times 10^{-11}$ . Finally Davier
\cite{g-2-davier2003} reported a value based on \epemt data
$a_\mu^{had}=(6847\pm 60_{exp}\pm 36_{th})\times 10^{-11}$ as well
as another one based on $\tau^+\tau^-$ data, $a_\mu^{had}=(7090\pm
47\pm 12_{exp}\pm 38_{SU(2)})\times 10^{-11}$. Since these two
values are slightly inconsistent we will  first  consider a limit
coming from an average of the hadronic estimates from $\epem$
alone
as well as a more conservative limit encompassing both $\epem$ and
$\tau$ results.
\footnote{ The new update analysis of Davier et al.,\cite{g-2-davier2003} finds a similar $\tau$-based result
and a much better agreement between the $\epem$-based and
$\tau$-based value. We have not used these values in our analysis
as our conservative range encompasses already the newest estimates.}

The theoretical estimate is obtained after adding the pure QED contribution,
$a_\mu^{QED}$, the weak contribution, $a_\mu^{weak}$ and the three different
hadronic contributions: $a_\mu^{\rm had},a_\mu^{\rm LBL}$ and the NLO
hadronic contribution, $a_\mu^{\rm had/NLO}$,
$$a_\mu^{\rm theo.}=a_\mu^{\rm QED}+a_\mu^{\rm
weak}+a_\mu^{\rm had}+a_\mu^{\rm LBL}+a_\mu^{\rm had/NLO}.$$

With  the latest experimental data on the $g-2$
measurement \cite{g-2ex}, bringing the world average to
\beqn
a_\mu^{\rm exp.}=11659203 \pm 8 \times 10^{-10}
\eeqn
we get, after  averaging of the three different hadronic
contributions from $e^+e^-$ data alone,
\beqn
\delta a_\mu=a_\mu^{\rm exp.} - a_\mu^{\rm theo.}=\left(33.6\pm
8_{|{\rm exp.}} \pm 11.5_{|{\rm theo.}}\right) \;\times 10^{-10}
\eeqn
Adding linearly the theoretical error to a $3\sigma$ experimental
error leads to the allowed range,
\beqn
-2 \;<\; \delta a_\mu\;\times 10^{10} \;<\; 69 \label{3sigma_1}
\eeqn

Including the $\tau$ data analysis of Davier \cite{g-2-davier2003}
in the average, reduces the discrepancy between theory and
experiment
\beqn
\delta a_\mu=\left(27.8\pm 8_{|{\rm exp.}} \pm 11.3_{|{\rm
theo.}}\right) \;\times 10^{-10}
\eeqn
The $3\sigma$ allowed range now reads,
\beqn
-7.5 \;<\; \delta a_\mu\;\times 10^{10} \;<\; 63 \label{3sigma_2}
\eeqn
As the results have changed frequently in the last months and
since many issues need to be clarified in particular in the
estimation of the hadronic contribution, the most conservative
allowed range is found by including two allowed bands, instead of
performing an average. Combining the value obtained from $\epem$
data
 (Eq.~\ref{3sigma_1})
 with the one from $\tau$ data alone of Davier, one gets a
 $3\sigma$ range
\beqn
-25 \;<\; \delta a_\mu\;\times 10^{10} \;<\; 69 \label{3sigma_3}
\eeqn
where the lower bound is set by the $\tau$ data and the upper
bound from Eq.~\ref{3sigma_1}. We will refer to Eq.~\ref{3sigma_3}
as our conservative bound on \gmuon.
 \footnote{ This last bound is similar to the one we had
obtained (at $2\sigma$)\cite{nous-susy02}
 before the new
calculations and the more precise results came out in summer 2002
\cite{g-2ex}.}

In the MSSM, $a_\mu$ gets a contribution both from neutralino and
chargino loops, the latter being dominant. One expects large
effects for large $\tan\beta$  and light sfermions
($\tilde{\mu}$/$\tilde\nu_\mu$) \cite{g-2susy}. Typically,  the
sign of $\mu$ is strongly correlated to the one of $\delta a_\mu$.
However there are special cases where
 cancellations can occur between
 the chargino and the neutralino diagrams thus flipping  the relative sign
of $\delta a_\mu$ and $\mu$. This happens for  charginos much
heavier than  neutralinos
 and the resulting value of $\delta a_\mu$ is small. This property will open a small corner
 of parameter space where models with $\mu<0$ will be allowed  even
 when imposing the bound in Eq.~\ref{3sigma_1}.
 When $\mu>0$ on the other hand, one expects rather mild constraints on the
 parameters of the MSSM, furthermore the constraints will be mostly
 in the chargino/slepton sector rather than directly on the neutralino sector.

\subsection{$\bsgamma$}

Our calculation of the $b\ra s\gamma$ closely follows the approach
of Kagan and Neubert \cite{bsgKagan}(see also \cite{bsgCMM}) with
NLO, bremstrahlung and some non-perturbative  effects taken into
account. This approach makes it easier to include the effects of
New Physics. Our rates refer to the so-called ``faked" total rate
with a cut on the photon energy corresponding to $\delta=0.9$
($E_\gamma
> m_b/20$). We have updated this analysis by including in a simple
way the recent suggestion of Misiak and Gambino
\cite{bsgGambino-Misiak} of using the $\overline{MS}$ charm mass.
The Wilson coefficients for the standard model and the charged
Higgs are evaluated at the NLO \cite{bsgCDGG-smh}, whereas we only
include the remaining SUSY contributions at LO \cite{bsgCDGG-susy}
with however the important inclusion of the enhanced  large $\tgb$
effects \cite{bsgDGG} (SUSY threshold corrections to the running
$b$ mass). For the latter we include both the strong $\alpha_s$
contribution as well as the Yukawa contribution for the SUSY
Wilson coefficients as well as the charged Higgs and the Goldstone
contribution\footnote{We have corrected some typos in
\cite{bsgDGG}. We thank Paolo Gambino for checking our results and
agreeing with our implementation.}.  Our standard model value
(with scale parameters set at $m_b$), gives $Br(b\ra s\gamma)=3.72
\;\times 10^{-4}$ while the scale and other parameter uncertainty
($\alpha_s$, CKM matrix elements) are about $10\%$. To bound the
SUSY contribution we take the world weighted average of the {\tt
CLEO} \cite{bsgCLEO} {\tt BELLE}\cite{bsgBELLE} and {\tt ALEPH}
\cite{bsgALEPH} measurements
\beqn
\label{bsgexp} Br(b\ra s\gamma)=3.23\pm .42 \;\times 10^{-4}
\eeqn

We require that after allowing for the (scale) uncertainty in the
theory calculation the result must be within $2\sigma$ of the
experimental result, Eq.~\ref{bsgexp}. Since the theory
uncertainty is roughly constant over the SUSY parameter space and
in order to have a faster scan we have allowed for  a conservative
fixed uncertainty of $10\%$ independently of the SUSY parameters.
Thus in effect we require the theory prediction to fall within the
range
\beqn
\label{bsgbound} 2.04 < Br(b\ra s\gamma)\;\times 10^{-4} < 4.42
\eeqn
In the MSSM,  contributions to the $b\ra s\gamma$ depends mostly
on the squark and gaugino/Higgsino sector as well as on the
charged Higgs. The heavy squarks that we consider do not
completely decouple and one can get substantial corrections to the
SM branching ratio. In particular at large $\tan\beta$ there is a
strong $A_t$ dependence through a term $\propto  A_t \tan\beta$
from the mixing in the stop sector. However, the light Higgs mass
is also sensitive to the  mixing in the stop sector. As long as the pseudoscalar is heavy, one finds
that for small values of $\tan\beta$, one can find
 values of $A_t$, typically $A_t$ large and positive, that satisfy both the
 Higgs mass  limit as well as the
$\bsgamma$ bound. For light pseudoscalars on the other hand we will see that
 it is very difficult to  satisfy the $\bsgamma$ bound.
 At large values of $\tan\beta$, the Higgs bound
is more easily satisfied and one can pick a set of values
for $A_t$ (typically $A_t<0$) that allows the correct amount of
$\bsgamma$ whether or not the pseudoscalar is heavy. Note that the light squarks contribution  to
$\bsgamma$ rapidly becomes too important. This again justifies our
choice of large squark masses.

In the end we find that because of the free mixing  parameter in the stop sector,
the  $\bsgamma$ constraint has a mild impact on the
 models considered at least as concerns the mass of the neutralino.
 One exception is the light pseudoscalar case
 that will be discussed in more details in the last section.

\subsection{$\bsmu$}
The {\tt CDF}  experiment at {\tt Fermilab} has obtained an upper
bound on the branching ratio B.R.($\bsmu<2.6\times 10^{-6}$)
\cite{CDFbsmumu} and should be able to reach B.R.($\bsmu<2.\times
10^{-7}$) in RunIIa. In the SM, this branching ratio is expected
to be very small ($\approx 3\times 10^{-9}$). In the MSSM SUSY
loop contributions due to chargino, sneutrino, stop and Higgs
exchange can significantly increase this branching ratio. In
particular, the amplitude for Higgs mediated decays goes as
$\tan\beta^3$ and orders of magnitude increase above the SM value
are expected for large $\tan\beta$. This process is relevant
mainly in the light pseudoscalar scenario. Our calculation is
based on \cite{bsmumubobeth} and agrees with \cite{bsmumudreiner}.
$\Delta m_b$ effect relevant for high $\tan \beta$ are taken into
account.

\subsection{$Z \ra b\bar b$}
We have also included the constraint from $Z \ra b\bar b$ although
in most cases it is completely harmless. For example, the charged
Higgs also contributes to $Z\ra b\bar{b}$. However this puts a
constraint on the Higgs sector only
 for Higgs masses below the {\tt LEP} limits. Our calculation of $Z \ra b\bar b$
is along the lines of \cite{zbb-higgs+} but we have corrected a
few typos contained in \cite{zbb-higgs+}.

\section{The lower bound on $\mneuto$ when $\ma=1$~TeV}

In this section we present our results for the case of a heavy pseudoscalar
after imposing all direct
and indirect constraints. In particular we have imposed only the
conservative limit on \gmuon, Eq.~\ref{3sigma_1}. Nevertheless it is worth
discussing the impact of the much stricter bound, Eq.~\ref{3sigma_3}. As
the allowed region in $\mu<0$ models changes drastically depending
on how one evaluates the hadronic contribution, we will defer the
extensive discussion of this constraint in the section devoted to
the $\mu<0$ case.
\subsection{$\mu>0$}
We first consider the case $\mu>0$. In these scenarios, the relic
density of dark matter provides the main constraint on models with
light neutralinos. Indeed the {\tt LEP} limits on charginos and
neutralinos allow very low values for neutralino masses provided
$M_1<<M_2$. The \gmuon~ constraint is in general easily satisfied
except for very large values of $\tan\beta$. The constraint for
$\bsgamma$ also affects mostly models with large values of
$\tan\beta$.

\noi 1) Heavy sleptons: $\m0=500$GeV.

\begin{figure*}[bhtp]
\begin{center}
\vspace{-1.3cm}
\includegraphics[width=16cm,height=10cm]{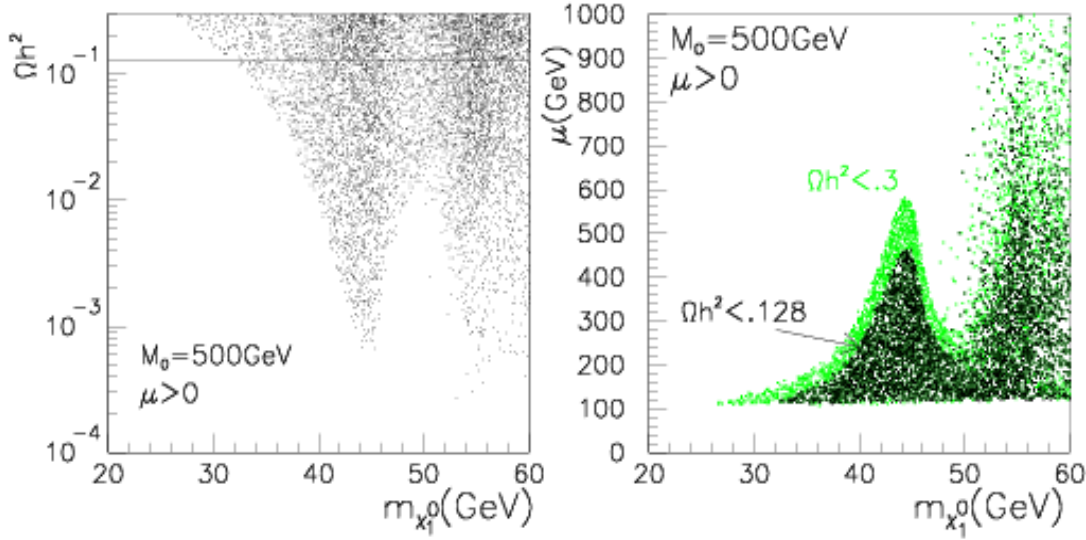}
\vspace{-2.cm} \caption{\label{m0500}{\em a) Relic density of the
LSP vs $\mneuto$ in the MSSM with $\m0=500$GeV.  To guide the eye
the {\tt WMAP} upper bound is displayed. b) Values of $\mu$
consistent with the {\tt LEP} and relic density constraints.
 We scan on $\tan\beta,M_2,\r12$ and $\mu$.
 }}
\end{center}
\end{figure*}
In this situation all sfermions are heavy. As a consequence their
contribution to the relic density is negligible. Annihilation of
the LSP dark matter through $s$-channel $Z$ and Higgs exchange is
on the other hand possible. Then in order to have a sufficient
neutralino annihilation rate, to satisfy the upper limit on the
relic density, a large enough coupling to the $Z$ is necessary.
This requires  the LSP (which is mostly a bino) to have a certain
amount of Higgsino component, especially as one moves away from
the $Z$ peak. Bino-Higgsino mixing has little \tgbt dependence and
scales like $M_Z/\mu$ which calls for the smallest $|\mu|$
possible. It can also be enhanced somewhat if there is little
splitting between $M_1$ and $\mu$, but this would generally not
minimise the LSP mass. We scanned over the parameters
$\r12,M_2,\mu$ and $\tan\beta$ as specified in Eq.~\ref{scan}. The
relic density as a function of the neutralino mass clearly shows
the effect of the $Z$ peak and the lower limit on the neutralino
mass, $\mneuto \approx 32 GeV$, Fig.~\ref{m0500}a. Note that
although some points give a relic density that is much too low,
models with $\Omega h^2\approx 0.1$ can be found for any  value of
the neutralino mass consistent with the lower limit of $\approx 32
GeV$.
\begin{figure*}[hbtp]
\begin{center}
\includegraphics[width=14cm,height=9cm]{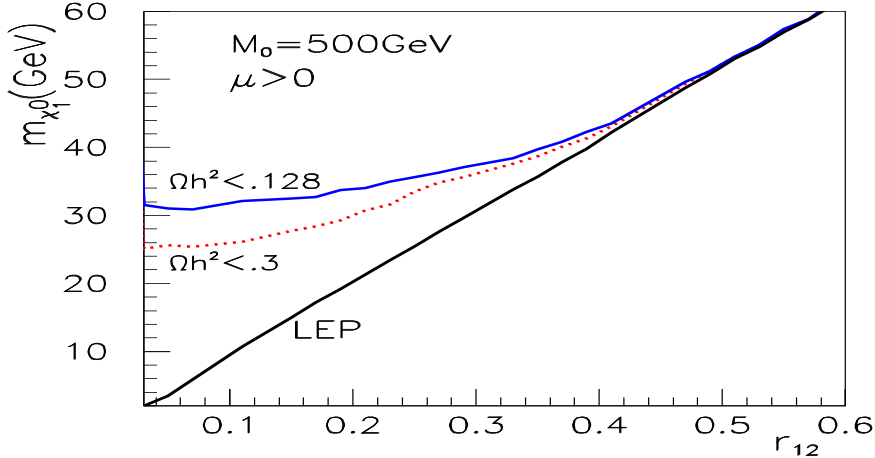}
\vspace{-1.7cm} \caption{\label{m0500_min}{\em Lower limit on
$\mneuto$ vs $\r12$ after imposing the relic density constraints.
 The limit from LEP2 alone is also displayed. Here $\m0=500$~GeV, $\mu>0$ and
 we scan on $M_2,\tan\beta,|\mu|$. The region below the lines is excluded.
 }}
\vspace{-.5cm}
\end{center}
\end{figure*}
In particular for the lightest neutralinos in this scenario (say
within $2$GeV of the lower limit) the value of the relic density
falls within the range of {\tt WMAP}. Incidentally,
Fig.~\ref{m0500} also shows that the relic density can drop quite
dramatically even beyond the $Z$-peak and upon inspection even
higher $\mu$ values are allowed in this range. This corresponds to
the contribution to the $s$-channel lightest Higgs $h$ with
$\mneuto\approx \mh/2$. Since in the scenario with $M_A=1$TeV,
$\mh>113$~GeV this occurs beyond the $Z$-peak. One of the reasons
this second peak looks broader is because we are scanning over a
wide range of $h$ masses. Another reason is that the coupling of
the (almost) bino LSP is larger for the Higgs than it is for the
$Z$. For the latter the coupling is in fact quadratic in the
Higgsino-bino mixing. As we will discuss later, we expect the
presence of light Higgses other than $h$ to reduce the value of
$\Omega h^2$ and for masses of the order of the $Z$ mass to allow
an even lower limit of neutralino LSP.\\
\noi Fig.~\ref{m0500}b clearly shows the preference for a
significant Higgsino component, the small $\mu$ region, when the
neutralino mass is below $M_Z/2$. The minimum value for the LSP
mass, $\mneuto \approx 32 GeV$,  occurs for $\r12<.2$, see
Fig.~\ref{m0500_min}. As $r_{12}$ approaches the mSUGRA value,
$\r12=0.5$, the limit from the chargino mass at {\tt LEP}
dominates and the relic density constraint has no effect. This
lower bound is more or less independent of $\tan\beta$ in the
range under consideration $5<\tan\beta<50$. Fig.~\ref{m0500_min}
also shows how the improvement of the bound on the relic density
from {\tt WMAP} has strengthened the lower limit on $\mneuto$  by
$\approx 5 GeV$ when $\r12<.2$. For larger values of the non
universality parameter, the improvement from {\tt WMAP} is only
marginal. Indeed, already with  $\Omega h^2<0.3$, one is confined
to a region
 ($\mneuto>35$~GeV ) close enough to the $Z$ peak so that $\Omega h^2$ drops sharply with $\mneuto$ (Fig.~\ref{m0500}~a).

The preference for the small $\mu$ region
 also implies  that  light neutralinos, say
below 40GeV, are necessarily accompanied by light  charginos
(below $250$GeV). However, as soon as one moves closer to $M_Z/2$,
the Higgsino component does not have to be too small and one
generally  has $\mchargo < 450$GeV. This has important
consequences for colliders. For example it could be difficult for
the {\tt Tevatron}   to find charginos in the trilepton channel
($\chi^+\chi_2^0$).
 Similarly  a 500GeV collider would not be able
to  find charginos in some scenarios as will be discussed in the
last section.
Note that under the conditions specified here, no improvement on
the lower limit on the neutralino can be set from \gmuon.

\noi 2) Light sleptons: $\m0<500$GeV
\begin{figure*}[bhtp]
\begin{center}
\includegraphics[width=16cm,height=8cm]{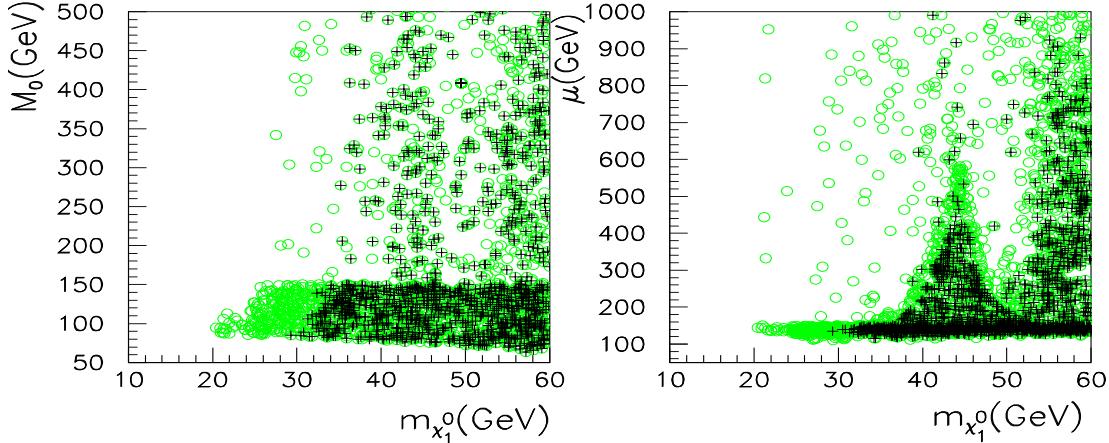}
\vspace{-2cm} \caption{\label{t10r1+}{\em Region allowed after
imposing {\tt LEP} and relic density constraints
  for $\tan\beta=10$, $\r12=0.1,\mu>0$ and $\ma=1$TeV,  a) in the $\m0-\mneuto$ plane b) in the $\mu-\m0$ plane. Here we scan on $\m0,M_2,\mu$ and $A_t$.
  Circles (crosses) have   $\Omega h^2 < 0.3 (0.128)$.
Half the points in the scan are generated in the region
corresponding to $\m0,\mu<150$~GeV. }}
\vspace{-.2cm}
\end{center}
\end{figure*}

In addition to the effect of the $Z$ exchange one expects the
contribution from $t$-channel sfermions to $\chi\chi\ra f\bar{f}$
to weaken the constraint from the relic density. Here the
constraints on the masses of selectrons/smuons and staus from LEP2
are important. The latter is relevant for large values of
$\tan\beta$ when the $\tilde\tau$ can be much lighter than the
selectron. As the left-right  mixing in the stau sector is
proportional to $A_\tau-\mu \tan\beta$ this  is especially true in
the large $\mu$ region. The (lightest) $\stau$ contribution to the
relic density in setting the lower bound on the LSP is largest
among all sleptons. Not only there is a $S$-wave contribution
($\propto m_\tau^2$) but the (dominant) $P$-wave contribution also
features an additional (positive) left-right mixing. But most
importantly this mixing allows for the smallest mass of the
$\stau$ compatible with the lowest mass on the $\stau_1$ set by
{\tt LEP} which is about $10$GeV lower than for the lightest
smuon and selectron.

First consider the case  $\tan\beta=10$ and $\r12=0.1$.
With the constraint $\Omega h^2<0.3$, one finds that at
low values of  $\m0$, the parameter that governs the mass of the right-handed sfermions,
 the lower limit on $\mneuto$ goes down  by a few GeV's relative to the one
 at large  values of $\m0$, Fig.~\ref{t10r1+}. However  the tail at low $\m0$
 is cut-off to a large extent (by $\approx 10$GeV)
 when taking into account {\tt WMAP}.
When imposing the constraint
$\Omega h^2<0.3$, we  find that one does not rely exclusively on the Higgsino
content of the neutralino to derive a lower bound.
 Large
$\mu$ values are compatible with neutralinos $\mneuto\approx
18$GeV . However, with the much tighter constraint from {\tt
WMAP},  and after including the {\tt LEP} constraint, the
$t$-channel sfermion exchange alone is not sufficient to bring the
relic density in the allowed range for $\mneuto<35$GeV. Once again
a non-negligible Higgsino component is required to have sufficient
coupling to the $Z$ (Fig.~\ref{t10r1+}). This entails, as was the
case for heavy sleptons, that  light neutralinos with masses close
to the lower bound are necessarily accompanied by light charginos.
However, for $\mneuto\approx M_Z/2$,  $\mchargo$ can now reach as
much as $\approx 1$TeV.

\begin{figure*}[hbtp]
\begin{center}
\includegraphics[width=16cm,height=11cm]{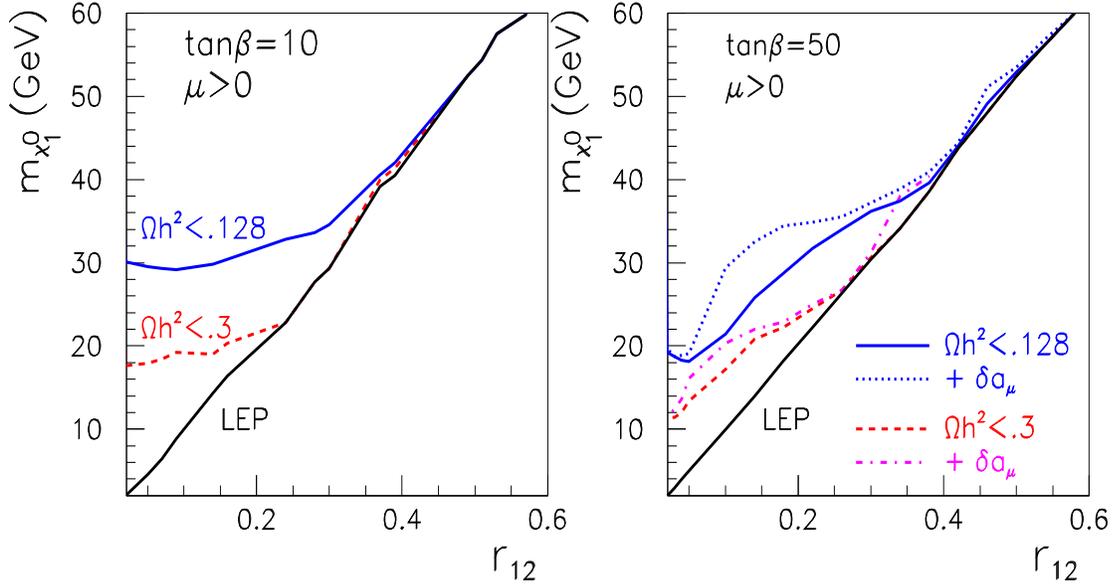}
\vspace{-2cm} \caption{\label{t10t50}{\em  Lower bound on the
neutralino mass vs $\r12=M_1/M_2$ for $\mu>0$ and a) $
\tan\beta=10$, b) $\tan\beta=50$. All direct and precision
measurements constraints are combined with the {\tt LEP} limits.
We scan over $M_2,\mu,\m0$ and $A_t$.}}
\end{center}
\end{figure*}
We show in Fig.~\ref{t10t50} the lower limit on the neutralino mass
for $\tan\beta=10$ and $\tan\beta=50$ as a function of $r_{12}$
after scanning over $M_2,\mu,\m0$ and $A_t$. The relic density
improves the constraint on the mass of the light neutralino only
after having taken into account the LEP2 constraint  on selectrons
(and on staus for large values of $\tan\beta$) as well as the
constraint from $\epem\ra \neuto\neutt,\neuto\neutth$. After {\tt
WMAP}, the absolute lower limit moves to $\mneuto\approx 29$GeV
for $\tan\beta=10$. This limit is obtained in the range $\r12<0.1$
and increases  with $\r12$. In the whole region $\r12<0.3$, the new
{\tt WMAP} data improves by up to $10$GeV the lower limit on the
neutralino mass obtained with pre-{\tt WMAP} data.
 As for heavy sleptons, the constraint on the
chargino mass from LEP2 sets the lower limit on the neutralino
mass when $0.3<r_{12}<0.6$. We have here also applied the upper
bound on the \gmuon~, Eq.~\ref{3sigma_3}. However this additional
constraint does not affect  the lower bound on the neutralino mass. For
$\tan\beta=50$, a lower limit of $\mneuto>18$GeV is obtained for
$\r12<.06$ and increases rapidly with $\r12$.
\footnote{In \cite{Plehn-lsp} a lower limit of 18 GeV was obtained
for any value of $\tan\beta$, using $\Omega h^2<0.3$, however a
universal limit on $m_{\tilde{l}}>100$GeV was imposed.}
 This lower
limit lies below the one obtained for lower values of $\tan\beta$.
The main reason is  the contribution of the channel
$\neuto\neuto\ra\tau^+\tau^-$ with $\tilde{\tau}$ exchange which
is enhanced because of the large $\stau_L$-$\stau_R$ mixing. The
new {\tt WMAP} data sets the lower limit on neutralino as long as
$\r12<.4$. Although we have taken a unified scheme for generating
the slepton masses along Eq.~\ref{m0running} with a common $M_0$
at high scale, we would have arrived at the same LSP lower limit
had we varied the slepton masses independently. This is due, as
explained earlier, to the fact that from {\tt LEP} the lowest
bound on the slepton applies to the $\stau$. On the other hand to
use the bound from \gmuon~ to further constrain the parameter space
tacitly assumes universality, at least as implemented through
Eq.~\ref{m0running}. Fig.~\ref{t10t50}b) shows the impact of
applying the \gmuon~ bound alongside the relic constraint. For
$\tgb=10$ there is no impact, however for $\tgb=50$ although the
lowest bound at very small $r_{12}$ is hardly affected by \gmuon,
for intermediate $r_{12}\sim 0.1-0.2$ one can improve the limit on
the LSP bound by about $6$GeV. This is related to the fact that
both the smallest values of $\m0$ as well as of $M_2$,
corresponding to light smuons and charginos,  give too large a
contribution to \gmuon~ as will be discussed below.

\subsection{$\mu<0$}
It has often been claimed that models with negative values of
$\mu$ are ruled out by the \gmuon. Within our conservative
approach this is not the case not only because the present data
and the calculation of the hadronic contribution  allows $\amu<0$
 but also because with nonuniversal gaugino
masses the sign of $\amu$ and $\mu$ are not necessarily
correlated. Considering the potential relevance of the \gmuon~ for
$\mu<0$ models, we will first discuss only this constraint
together with {\tt LEP} before folding in the constraints from the
relic density and $\bsgamma$. To a certain extent we will see that
these two sets of constraints work in the same direction
restricting the region with light sleptons and light
charginos/neutralinos.

\subsubsection{\gmuon}
\begin{figure*}[htbp]
\begin{center}
\vspace{-1cm}
\includegraphics[width=16cm,height=12cm]{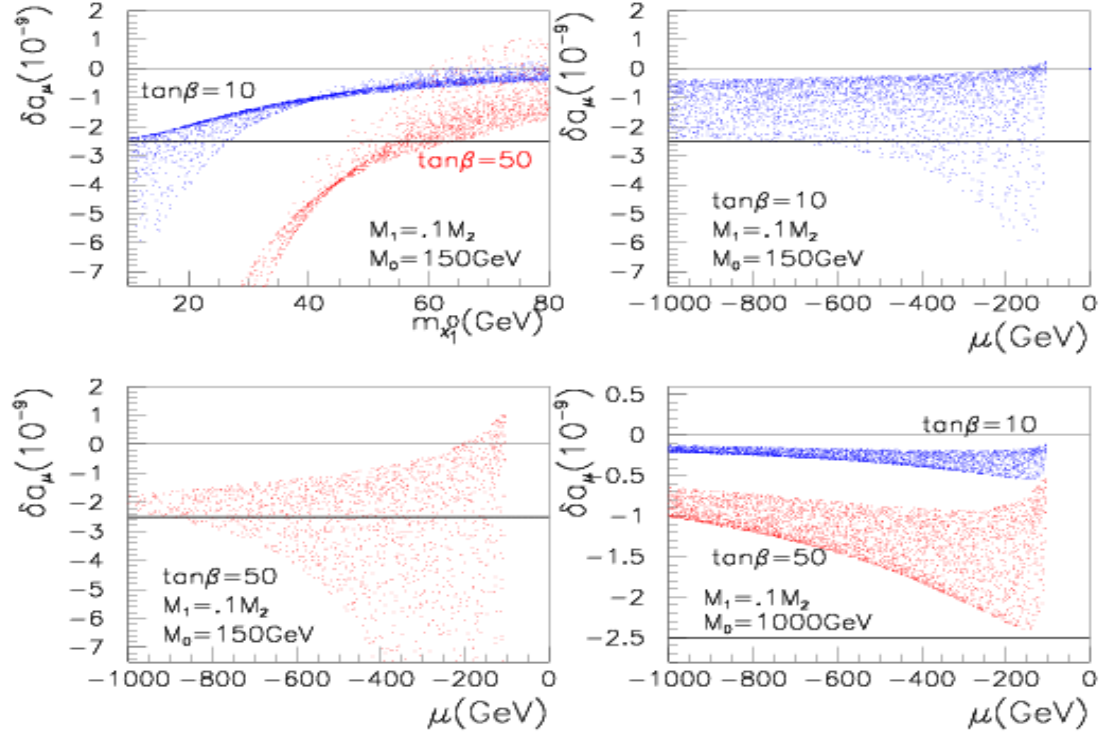}
\vspace{-1cm} \caption{\label{g-2_mu}{\em  $\delta a_\mu$ in
models where $\mu<0$, $\r12=0.1$ with $\tan\beta=10,50$ and  a-c)
$\m0=150$GeV d) $\m0=1000$GeV. The constraint on $\amu$ selects
all models above the thick line. Here only {\tt LEP} constraints
are imposed.}}
\end{center}
\end{figure*}

The most important contribution to the \gmuon~ occurs for light
smuons and light charginos and is enhanced at large values of $\tan\beta$.
First consider the case  $\r12=0.1$ and light sleptons,  $\m0=150$
GeV. When $\tan\beta=10$,  only a few models with very light
neutralinos exceed the  conservative limit Eq.~\ref{3sigma_3}, see
Fig.~\ref{g-2_mu}a.
However, as $\tan\beta$ becomes large, one derives a lower limit
on the neutralino of $\mneuto>50$GeV for $\tan\beta=50$. It is in
the small $|\mu|$ region that one finds the largest deviation in
the value of $\delta a_\mu$. For $\tan\beta=50$ one quickly
exceeds even the  conservative $3\sigma$ bound. However a
cancellation between the chargino and the neutralino diagrams can
change the relative sign of $\delta a_\mu$ and $\mu$,
Fig.~\ref{g-2_mu}b-c. As the neutralino contribution is in general
rather small, for this type of cancellation to occur,  the
chargino contribution must be somewhat suppressed (the large $M_2$
region) furthermore  a significant amount  of Higgsino/gaugino
mixing is required, (Fig.~\ref{g-2_mu}b(c)). Typically this sign
flip  is compatible with light neutralinos, say $< 50$GeV,  only
in the context of nonuniversal models  where the appropriate
hierarchy of parameters can be found, that is  $M_1<\mu<<M_2$.
For heavy sleptons, the predicted values for
$\delta a_\mu$ are in general much smaller and  all the
 parameter space (for $\r12=0.1$) satisfies  the experimental bound
 (see Fig.~\ref{g-2_mu}d).

\begin{figure*}[tbhp]
\begin{center}
\vspace{-1.2cm}
\includegraphics[width=16cm,height=14cm]{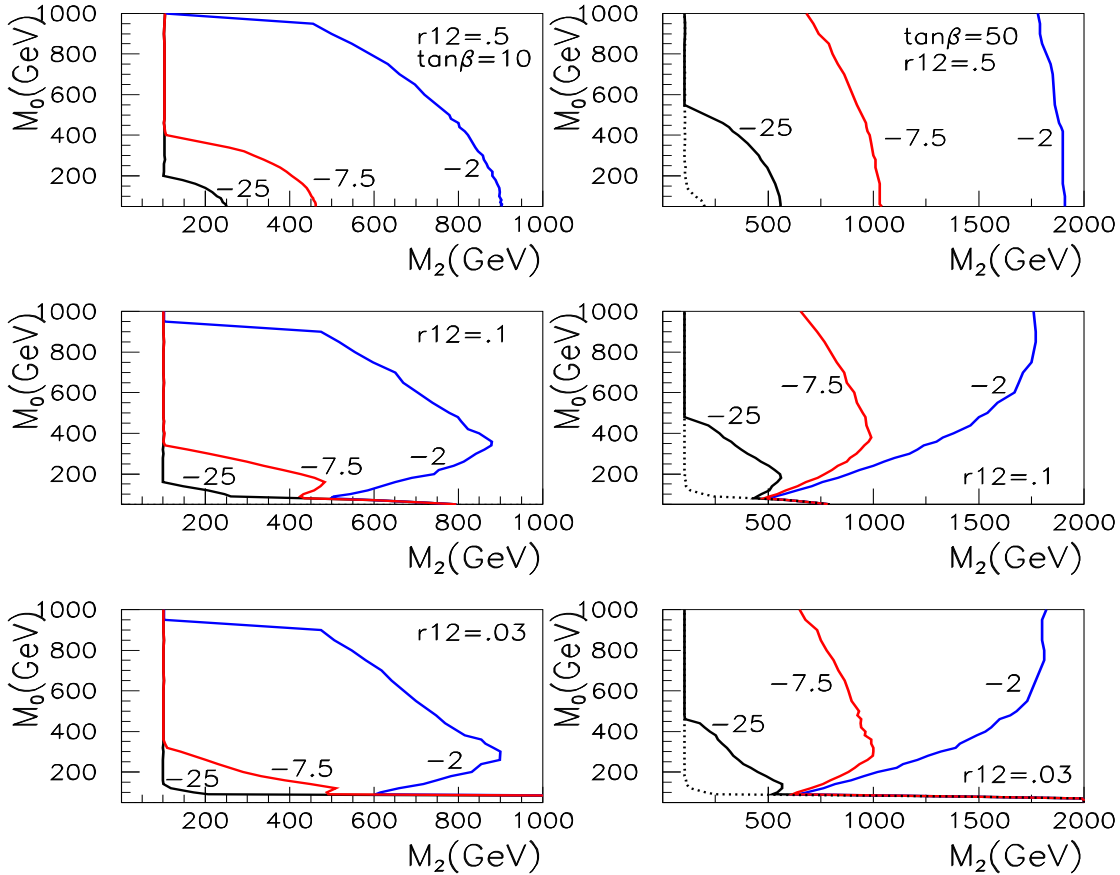}
\vspace{-1.9cm} \caption{\label{contour_g2}{\em  Contours $\delta
a_\mu\times 10^{-10}$
 in the $\m0-M_2$ plane  with $\mu<0$ for $\tan\beta=10$(left) and $\tan\beta=50$(right).
From top to bottom $\r12=0.5,0.1,0.03$. The top panel $\r12=0.5$ is
based on the usual GUT assumption. Only {\tt LEP} constraints are
folded in. For $\tan\beta=50$, the region allowed by $\bsgamma$
lies to the right of the dotted line and as can be seen it does
not further restrict the bound set by \gmuon. }} \vspace{-.2cm}
\end{center}
\end{figure*}

Fig.~\ref{contour_g2} shows the exclusion region in
 the $\m0-M_2$ plane for three different exclusion
values for $\delta a_\mu$ (depending on the assumptions on the
calculation of the hadronic contribution)  and for different
values of the non universality parameter. Here both $\mu$ and
$A_t$ are kept as free parameters and the {\tt LEP} limits on
$\mh$ and on charginos and sleptons are imposed. The \gmuon~ rules
out regions where $\m0$ and $M_2(M_1)$ are small. However, no
lower limit on the neutralino can be derived  as light
charginos/neutralinos are allowed when sleptons are very heavy. If
one would impose the more restrictive limits on $\amu$,
Eq.~\ref{3sigma_1}, it would be possible to accommodate light
sleptons and neutralinos in models where $\r12<.5$,
Fig.~\ref{contour_g2}. Indeed, the partial cancellation between the
chargino and neutralino diagrams  leads to  a limit on $M_2$ that
is not as strong as in the universal case. With this restrictive
bound it is possible to set a lower limit on the neutralino from
\gmuon~ only for large values of $\tan\beta$.

Note that we have only discussed explicitly the case $A_\mu=0$,
while corrections to the \gmuon~ are expected for large values of
$A_\mu$ it does not strongly affect the allowed region as long as
$A_\mu<1$~TeV. Indeed when $\tan\beta$ is small there is not much
constraint and when $\tan\beta$ is large we have in any case
$|\mu|\tan\beta >> |A_\mu|$.

\subsubsection{Combining all constraints}
\begin{figure*}[htbp]
\begin{center}
\includegraphics[width=16cm,height=11cm]{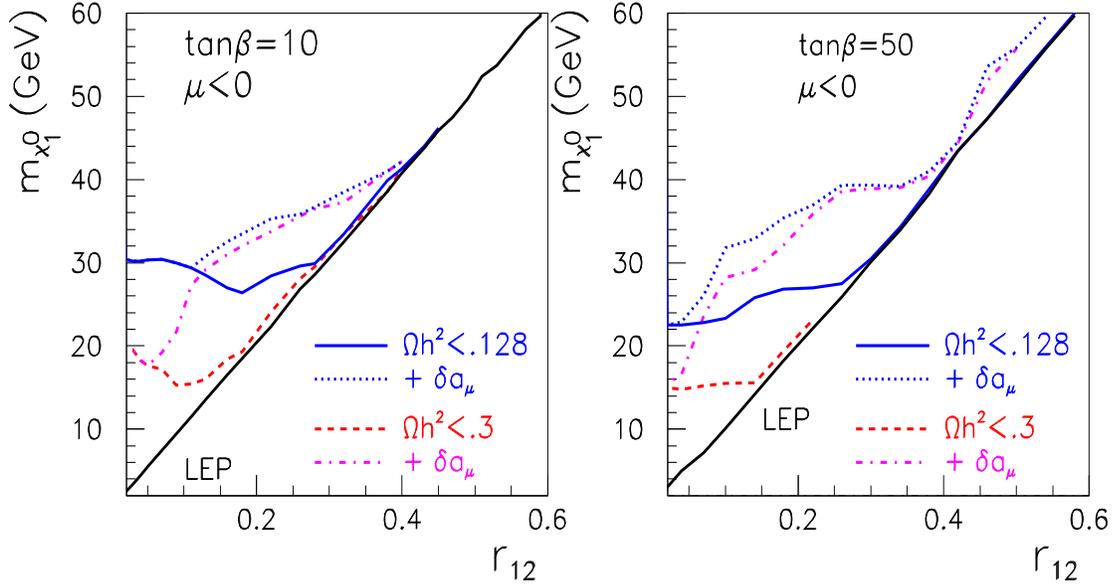}
\vspace{-2cm} \caption{\label{t10t50-}{\em  Lower bound on the
neutralino mass vs $\r12=M_1/M_2$ for $\mu<0$ and a)
$\tan\beta=10$, b) $\tan\beta=50$. All direct and precision
measurements constraints are combined with the relic density
limits (dotted lines) while the constraint from $\amu$ is removed
(full lines).}}
\end{center}
\end{figure*}
The impact of the relic density constraint coupled to direct
collider limits from {\tt LEP} is somewhat similar to what was
discussed in the case $\mu>0$. Here we stress the impact of adding
the indirect constraints from the \gmuon~.

\noi
{$\bullet$ \bf $\mu<0,\tan\beta=10$}\\
\noi
 With the new {\tt WMAP} data,
 it becomes increasingly difficult to satisfy the relic density constraint.
 Even with light sleptons, sufficient annihilation of neutralinos into fermion-antifermion pairs
requires the neutralino to be not too far away from the $Z$ pole.
As we had already found for $\mu$ positive the lower limit on the
neutralino mass increases significantly in the region $\r12<0.1$,
from $\mneuto>18$GeV based on the pre-{\tt WMAP} to
$\mneuto>29$GeV with {\tt WMAP}, Fig.~\ref{t10t50-}a. Note that in
this region, the bound on the $\epem \ra \tilde\chi_i\neuto$ cross
section  from {\tt LEP} has a significant impact in constraining
the small $\mu$ region. As a result, the lightest neutralino
allowed by  {\tt LEP} and {\tt WMAP} alone occurs for for larger
values of the non universality parameter, $\r12\approx 0.15$ with a
bound $\mneuto>26$GeV. It is only by imposing the \gmuon~ bound
that one increases the lower bound on the LSP in the region
$0.1<\r12<0.4$. It is also because of the importance of \gmuon~
that {\tt WMAP} does not sensibly improve the LSP bound in this
region compared to the pre-{\tt WMAP} data.

\begin{figure*}[htbp]
\begin{center}
\includegraphics[width=16cm,height=12cm]{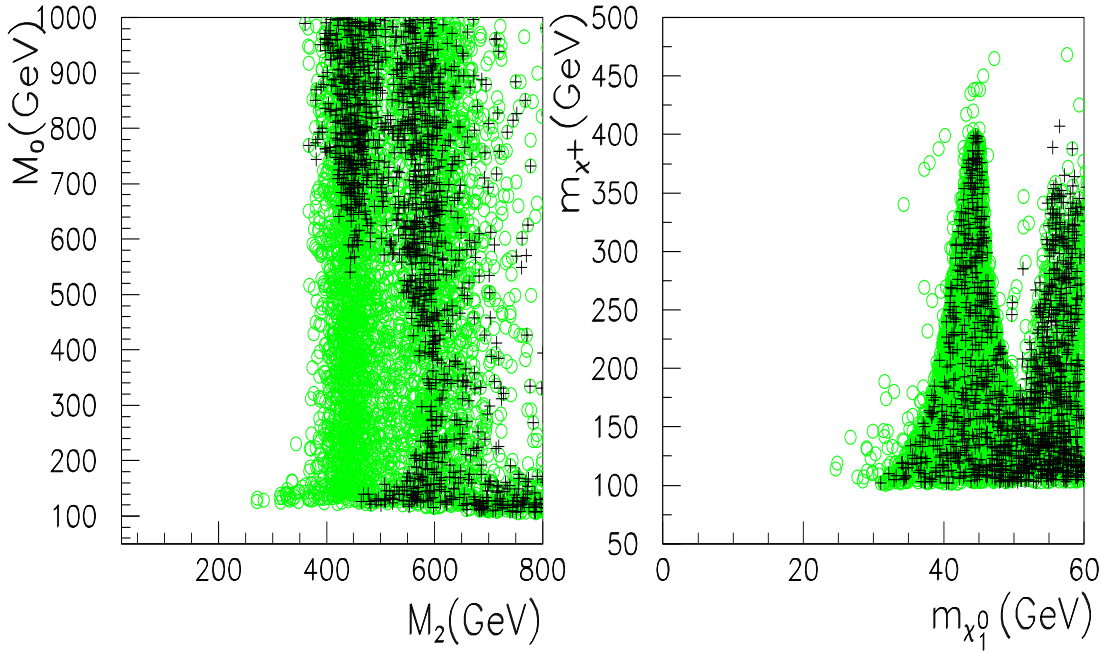}
\vspace{-2cm} \caption{\label{tg50snu}{\em Allowed region in the
$\m0-M_2$ plane using direct limits and {\tt WMAP} constraints
only (light grey circles) and when taking into account all
constraints (dark crosses). Here  $\tan\beta=50$, $\mu<0$ and
$\r12=0.1$.}}
\end{center}
\end{figure*}

\noi
{$\bullet$ \bf $\mu<0,\tan\beta>10$}\\
\noi For larger values of $\tan\beta$, even our most  conservative
bound on $\amu$ significantly   restricts the LSP mass, except for
the smallest $\r12$ values below about $0.03$. Past this value the
\gmuon~ bound is more restrictive in setting a limit on the LSP
mass, see Fig.~\ref{t10t50-}b and very much improves on the {\tt
WMAP} result. As a result the lowest LSP mass $\mneuto\sim22$~GeV
is found for $\r12\sim 0.03$. Had we used the pre-{\tt WMAP}
result, this limit would have been $6$GeV lower. The effect of
\gmuon~ for $\r12=0.1$ and $\tgb=50$ is more clearly seen in
Fig.~\ref{tg50snu}. For this set of parameters $\mneuto> 32$GeV
are allowed, an increase of nearly $10$~GeV had we not implemented
the \gmuon~ bound. Note that the \gmuon~ bound also cuts on the
region where both $M_2(M_1)$ and $M_0$ are smallest.
Fig.~\ref{tg50snu} also makes it clear that the lightest LSP have
mass near the $Z$ threshold which as we have argued calls for a
not too large $\mu$. In turn one sees that this scenario also
predicts charginos (with a large Higgsino component) that are
relatively light, for instance for $\mneuto<40$ GeV, one expects
$\mchargo<250$GeV.
 In the region $0.1<\r12 <0.3$, we find
that  combining all  constraints increases the lower bound on the
neutralino and that the increase is more significant for very
large $\tan\beta$. However, for this range of values for the non
universality parameter, {\tt WMAP} has much less impact. In this
region  the lower limit on the neutralino depends strongly on the
lower allowed value for $\amu$. Had we used the stricter bound on
\gmuon~, Eq.~\ref{3sigma_1}, we would not have found neutralinos
below $60$GeV for the large values of $\tan\beta$.

In summary we have found that the lower limit on the neutralino
mass is $\mneuto>29$GeV for $\tan\beta=10$ for either sign of
$\mu$ and $\mneuto>18(22)$GeV for $\tan\beta=50$ and
$\mu>0(\mu<0)$.
 The latter is allowed only for large mass splitting in the gaugino sector, $M_1 \ll M_2$. The muon anomalous magnetic moment  does
constrain models with  $\mu<0$ as well as the large $\tan\beta$ region except in the  region where $M_1\ll M_2$.
Furthermore, in the region $\r12> 0.1$ and when $\mu<0$
 the lower bound on $\mneuto$ strongly depends on
 how much negative one allows   $\amu $ to be.

\section{Lowering $\ma$}

\subsection{Limits on the LSP}

The effect of lowering $\ma$, within the range allowed by LEP, has
an impact on both the relic density constraint and the indirect
limits from the $B$ sector.
 The relic density can be lower than the one obtained in the heavy pseudoscalar case  due to the
increase of  the $s$-channel Higgs exchange contribution  (notably
the from the pseudoscalar $A$) to the neutralino annihilation
cross section. This had to be expected in view of  the importance
of the $s$-channel pole contribution as we have already seen for
the $Z$ and the lightest Higgs, $h$. Recall that in these models
$M_h$ and $\ma$ can be as low as $91.6$ GeV without being in
conflict with the LEP2 data. In the large $\tan\beta$ region, the
contribution of such light Higgses
 can be sufficient to bring the relic density below the upper bound even if one is quite far from the
$s$-channel pole  due to the enhanced coupling of the Higgs to $b$
quarks and $\tau$'s. This opens up a new region of parameter space
for light neutralinos ($\mneuto<18$GeV) as was pointed out
in \cite{Torino-lsp1,Torino-lsp2}.

\begin{figure*}[bhtp]
\begin{center}
\vspace{-1cm}
\includegraphics[width=16cm,height=10cm]{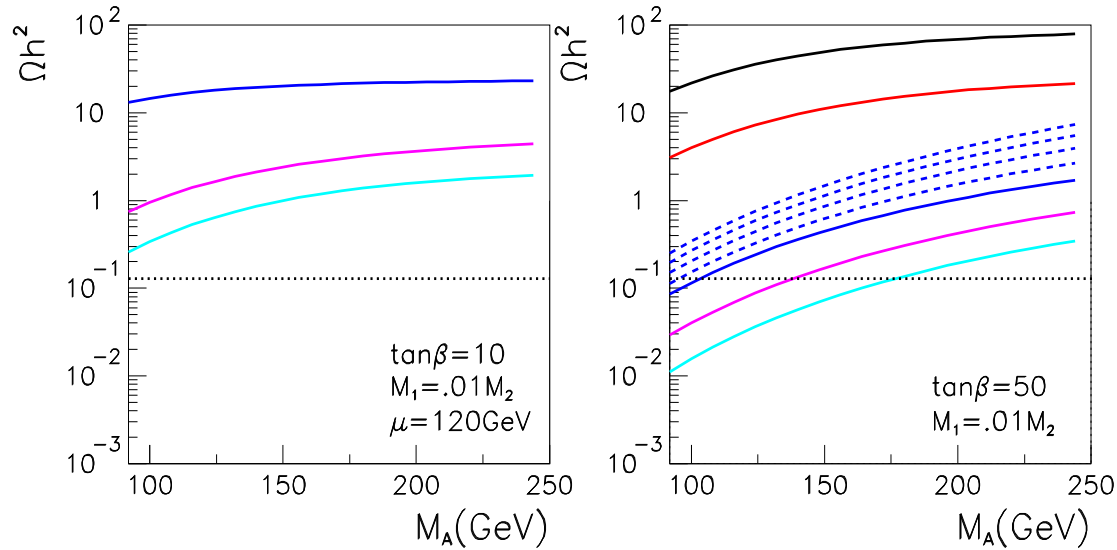}
\vspace{-1.2cm} \caption{\label{omega_ma}{\em  $\Omega h^2$ as a
function of $\ma$ for $\mu=120$GeV, $\m0=1000$GeV, $M_1=0.01M_2$
and a) $\tan\beta=10$ b) $\tan\beta=50$. Full lines correspond to
(from top to bottom) $M_2=800,1200,1800$ GeV (left) and
$M_2=300,500,800,1200,1800$ GeV (right). Additional dashed lines
show the effect of varying $\mu$ when $M_2=800$GeV (from  bottom
to top) $\mu=160,200,240,280$GeV. The dash line indicates the
current upper bound from {\tt WMAP}. Only direct {\tt LEP} limits
are imposed.}}
\end{center}
\end{figure*}

 The dependence of the relic density, for small $\mu$,  on
$\ma$ is displayed in Fig.~\ref{omega_ma}  for $\r12=0.01$ and for
$\tan\beta=10$ and $\tan\beta= 50$.
The upper bound on the relic density is satisfied most easily for
a light pseudoscalar, $\ma\approx 100-120$GeV and for large values
of $\tan\beta$. For instance $\tan\beta=10$ is totally excluded by
{\tt WMAP} for the set of parameters chosen, in particular 
for heavy sleptons, as seen in Fig.~\ref{omega_ma}a. Due to the
enhanced coupling of the pseudoscalar to the heavy fermions,
$\propto \tan\beta$, Fig.~\ref{omega_ma} also shows that changing
$\tgb$ from $10$ to $50$ makes the relic density compatible for
low values of $\ma$. The relic density decreases as one increases
$M_1$ (or $M_2$ since $\r12$ is fixed), thus increasing the LSP
mass. Once again, we confirm that in order to maximize the
Higgsino-bino mixing, small values of $\mu$ are preferred.
Generally,  the light neutralinos (say below 16GeV) are found when
$M_1<<\mu <<M_2$ and we therefore also expects relatively light
charginos as was the case with the heavy pseudoscalar scenario.

\begin{figure*}[htbp]
\begin{center}
\vspace{-1cm}
\includegraphics[width=12cm,height=10cm]{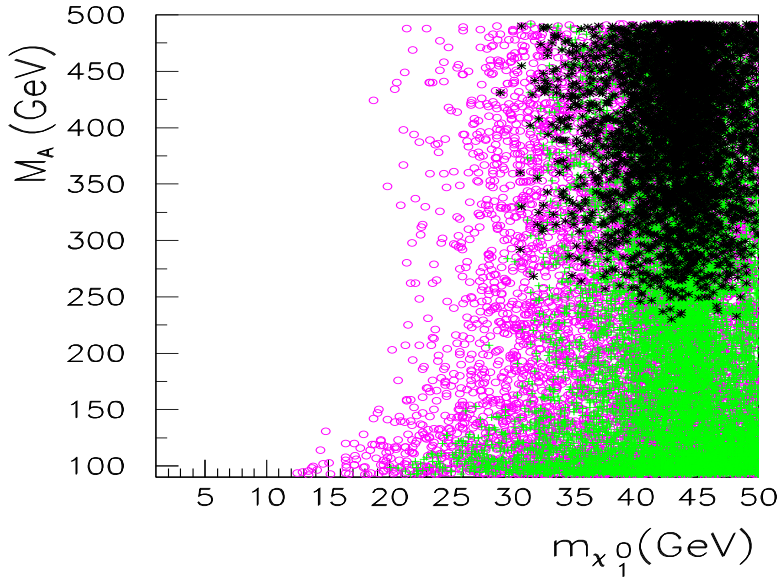}
\vspace{-1cm} \caption{\label{wmap_ma}{\em  Allowed values for
$\mneuto$ as a function of the pseudoscalar mass for
$\tan\beta=10$. The impact of the  {\tt WMAP} data on the relic
density is displayed: pink (light grey) circles corresponds to
$\Omega h^2< 0.3$ and green(medium grey) crosses to $\Omega h^2<
0.128$. The black crosses correspoind to the region  allowed after
including the $\bsgamma$ constraint. }}
\end{center}
\end{figure*}

\begin{figure*}[hbtp]
\begin{center}
\vspace{-1cm}
\includegraphics[width=12cm,height=9cm]{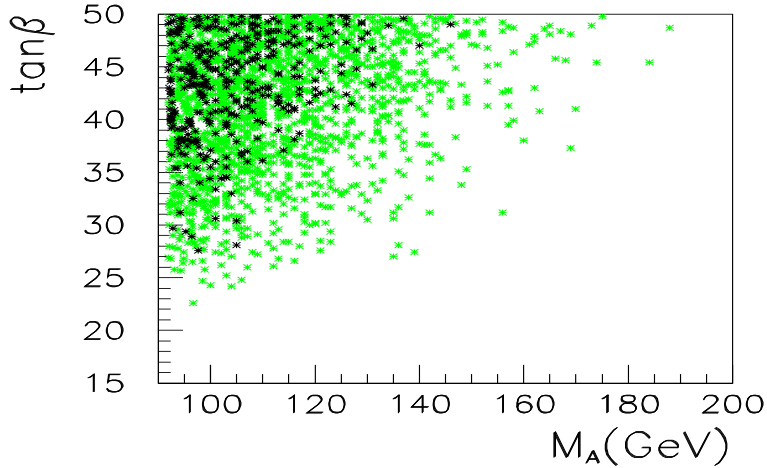}
\vspace{-1cm} \caption{\label{tbma}{\em  Region of the
$\tan\beta-M_A$ plane where one can find light neutralinos, Light
grey (green) crosses have $10$GeV$<\mneuto<16$~GeV while black crosses
correspond to $\mneuto<10$~GeV. Here all constraints are
implemented. }}
\end{center}
\end{figure*}

In models with low $\ma$ and large $\tgb$ one expects large
effects from the $b$ observables. Most affected are $\bsgamma$ and
$\bsmu$. On the other hand we have checked that the branching
ratio to $Z\ra b\bar{b}$ remains insensitive even in this
scenario. The branching ratio for $\bsgamma$  receives $\tan\beta$
enhanced contributions from both the charged Higgs diagrams as
well as the chargino/stop diagrams. Typically the contributions
form each types of diagrams individually could be far above the
allowed limit. However a partial cancellation between these two
contributions occurs  when $A_t$ is large and negative. For
example for $\tan\beta=50$, one needs values of $A_t<-750$~GeV.
For $\tan\beta=10$, on the other hand, even  the value
 $A_t=-2400$~GeV,
is not sufficient to induce a cancellation between the
  chargino sector and the large contribution from the charged Higgs
  to $\bsgamma$. This process then
forbids the very low values of $\ma$ for intermediate values of
$\tan\beta$, Fig.~\ref{wmap_ma}.
One sees for example that the lower bound for $\tan\beta=10$,
moves from $\mneuto\approx 12$GeV to $29$GeV after imposing the
$\bsgamma$ constraint. Models with very light neutralinos  that
are  compatible with the $B$-sector constraints as well as with
the relic density are then expected to be models with light
pseudoscalars and large values of $\tan\beta$ as one can see in
Fig.~\ref{tbma}.

\begin{figure*}[hbtp]
\begin{center}
\includegraphics[width=11cm,height=9cm]{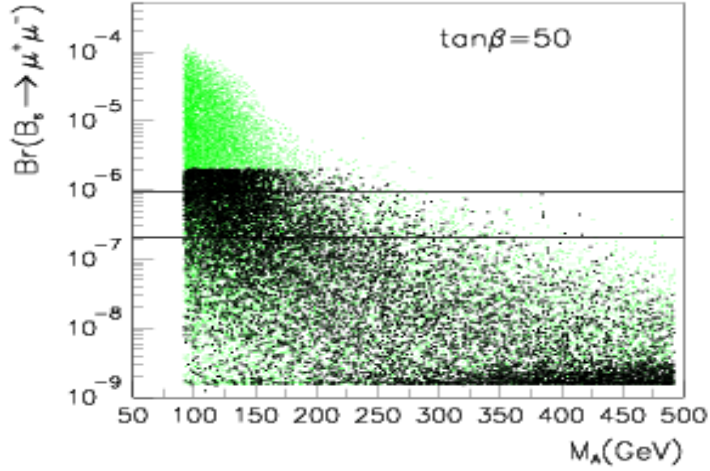}
\vspace{-0.5cm} \caption{\label{bsmu}{\em Branching fraction for
$\bsmu$ vs the pseudoscalar mass for $\tan\beta=50$, $\mu>0$. We
scan over all parameters and impose the {\tt LEP} and {\tt WMAP}
constraints (green/light grey) as well as all constraints (dark).
The preliminary  $95\%$ upper limit obtained by {\tt CDF}  with
113 $pb^{-1}$ ($9.5\times 10^{-7}$) and the expected reach of
RunIIa ($2\times 10^{-7}$) are also
displayed\cite{bsmu_leptonphoton}. }}
\end{center}
\end{figure*}
In these  models, the  branching ratio for $\bsmu$ can be strongly
enhanced at low values of $\ma$. In particular diagrams with
chargino loops could give a large contribution when the chargino
has a large Higgsino component. This corresponds to not too large
values of $M_2$. The predicted values for $\bsmu$ are displayed in
Fig.~\ref{bsmu} for a variety of models that pass both {\tt LEP}
and {\tt WMAP} constraints. One sees that the present limit from
the  {\tt Tevatron}  on the $\bsmu$ eliminates some models but
leaves open the possibility of light pseudoscalars.

\begin{figure*}[htbp]
\begin{center}
\vspace{-1cm}
\includegraphics[width=14cm,height=10cm]{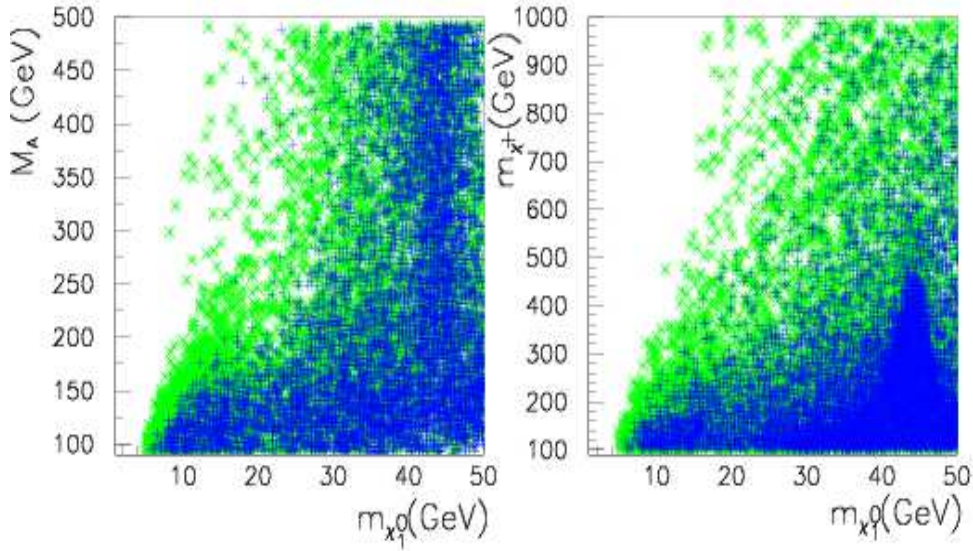}
\vspace{-1.5cm} \caption{\label{tan50_ma}{\em  Allowed values for
$\mneuto$ as a function of the pseudoscalar mass for
$\tan\beta=50$. The impact of the  data on the relic density is
displayed. Green stars (medium grey) correspond to $\Omega h^2<
0.3$ while black crosses correspond to $\Omega h^2< 0.128$. b)
Chargino masses as  a function of LSP mass. All constraints are
implemented. }}
\end{center}
\end{figure*}

A scan over the full parameter space, Eq.~\ref{scan}, with $92
{\rm GeV}<\ma<500$GeV and imposing all the above mentioned
constraints leads the lowest LSP mass bound of $\mneuto=6$~GeV.
\footnote{This agrees with the results of \cite{Torino-lsp2}.}
This LSP is found in models with large values of \tgbt
$\tan\beta>30$ and is associated with
 $\MH,\ma,|\mu| < 120$GeV. The sleptons need not be light.
The allowed region in the $\ma-\mneuto$ plane (Fig.~\ref{tan50_ma})
for $\tan\beta=50$ clearly shows that  when $\ma< 250$ GeV
 one can lower the  bound on the neutralino compared to the case
 when the pseudoscalar is $1$~TeV. Contrary to what we have seen for other cases,
 the impact of the {\tt WMAP} results is
 marginal as concerns the lower bound on the neutralino mass $\approx 6$~GeV, see
Fig.~\ref{tan50_ma}. Although  we found that a large number of
models were ruled out by the requirement of  $\bsgamma$
(Eq.~\ref{bsgexp}), this constraint  does not affect the absolute
lower limit on the neutralino mass, $\mneuto=6$~GeV, in the large
$\tan\beta$ scenario. Similarly the $\bsmu$ constraint does not
impact on this lower bound. Moreover, for  intermediate values of
$\tan\beta\sim 10-20$, we find that lowering $\ma$ does not affect
the lower limit on the neutralino mass which we derived for
$\ma=1$~TeV. This is due to an incompatibility with the $\bsgamma$
constraint as shown explicitly for $\tan\beta=10$ in
Fig.~\ref{wmap_ma}. In fact, restricting the analysis to the
region where $\mneuto<16$GeV, we found that only models with
$\tan\beta>25, |\mu|<400$GeV, $\ma<200$GeV and a large negative
stop mixing parameter $A_t<-750$~GeV were consistent with all
constraints. The first three conditions are necessary for
sufficient annihilation of the LSP into $\tau \bar \tau
(b\bar{b})$ pairs, while the last condition ensures cancellation
between the chargino and charged Higgs contributions to
$\bsgamma$. Obviously, as the coupling of the light Higgs scalar
is enhanced with $\tan\beta$, the range of Higgs masses allowed is
wider for larger values of $\tan\beta$, (see Fig.~\ref{tbma}).
With not so heavy pseudoscalar Higgs and  charginos, colliders,
and in particular the {\tt Tevatron},  might then have good
prospects to constrain the models or discover new particles in
such a framework as will be discussed next.

\subsection{Prospects at the {\tt Tevatron} }

Based on the existing {\tt Tevatron}
analyses\cite{TevatronHiggsWG2000}, we infer that the {\tt
Tevatron}  should be sensitive to some neutral Higgs bosons in a
large fraction of the models having very light neutralinos with
$\mneuto<16$~GeV. As we have just discussed, the various
constraints already impose that
 $\ma<220$~GeV and that  $\tan\beta$ be large
 to arrive at such low values for the neutralino mass (see Fig.~\ref{tbma}).
 Then the channel $p\bar{p}\ra b\bar{b} \phi$, with
$\phi\ra b\bar{b}$ for $\phi=h,H,A$ is rather sensitive to the
large $\tan\beta$ region due to the enhanced coupling to the $b$
quarks. For example, for $\tan\beta=50$, in the $M_{h_{max}}$ scenario, values of $\ma
\approx 105$GeV are excluded already with $0.1fb^{-1}$  of
integrated luminosity and with $2fb^{-1}$
 the region $\ma<200$GeV can be excluded at $3\sigma$ \cite{TevatronHiggsWG2000}.
 For $\tan\beta=40(30)$, $\ma<160(130)$ GeV can be excluded with $2fb^{-1}$.
For the very large values of $\tan\beta$
  under consideration  we have checked that the invisible mode, $h\ra\neuto\neuto$
carries only a small fraction of the total width so that
the $b\bar{b}$ branching ratio would not be significantly suppressed.
 The models with light Higgses, $M_{h,H,A}\approx 100$GeV   can also be probed at the {\tt Tevatron}
 in the charged Higgs channel
via disappearance searches for light charged Higgses  produced in
the decay of top quarks $t\ra H^+ b$. With $2fb^{-1}$ of
integrated luminosity in RunIIa, the region $m_{H^+}<130(150)$GeV
for $\tan\beta=30(50)$ can be excluded at $95\%$. However, as the
charged Higgs are heavier than the pseudoscalar, this channel does
not improve on the potential of the {\tt Tevatron} .

Moreover the required values of $\mu$ for models leading to
$\mneuto<16$GeV are such that the lightest chargino masses  can
also fall within the range accessible by the {\tt Tevatron} in
RunIIa. In particular, when  $\ma$ gets close to $200$GeV and
higher luminosities are needed for the Higgs channels, the
predicted values for the charginos cluster near $100-150$GeV.
Although a detailed analysis of chargino searches in the trilepton
mode in non-universal models has not been completed yet,
\cite{tri-leptons-nonuni},  it is expected that the {\tt Tevatron}
would be sensitive to charginos within this range.

Finally, the {\tt Tevatron}  can also probe the large
$\tan\beta$-light LSP scenario, via the $\bsmu$.
 In Fig.~\ref{bsmu}, we show the range of
predicted values for $\bsmu$ as well as the expected
sensitivity of RunIIa ($2\times 10^{-7}$). A large number of
models with light neutralinos,  predict a branching ratio above
this sensitivity. The largest branching ratios are found for $\ma<250$GeV.

To estimate the potential of RunIIa to probe models with very
light neutralino LSP, we assume the  expected limits,
$m_{A}<130(200)$~GeV, $m_{H^+}<130(150)$~GeV and $\bsmu<2\times
10^{-7}$, for values of $\tan\beta=30(50)$. For the chargino
channel we take as a guideline the value ${\mchargo}<150$~GeV.
Basically, we find that the additional region of parameter space
where light neutralinos were allowed as one lowered $\ma$ will be
probed entirely, at $\tan\beta=50$, by the {\tt Tevatron} , both
in the pseudoscalar searches as well as in the $\bsmu$. The
chargino searches will also somewhat help constraint the allowed
models. For $\tan\beta=30$, due to the lower reach in the Higgs
channel, we found some models with $\mneuto > 15 $GeV which could
escape detection in RunIIa.

In summary, decreasing the value of $\ma$ opens up new
possibilities for light neutralinos ($6-16$GeV). With Higgs and
chargino searches, the {\tt Tevatron}  should be able in the near
future to  cover the small remaining part of parameter space where
very light neutralinos ($\approx 6$GeV) can exist.

\subsection{Direct detection with light scalars}

The models we have just discussed admitting pseudoscalars (and
hence the other neutral Higgses) with masses close to the lowest
limit set by {\tt LEP} offer interesting prospects for the direct
detection experiments as well. The spin independent scalar cross
section for neutralino scattering on a nucleus is  sensitive to
squark and Higgs exchanges. However, in the models with heavy
squarks
 considered here,  the dominant contribution
 arises from  $t$-channel Higgs exchange.
 The dependence of the cross section
 on the mass of the
  Higgs scalars goes simply as $\sigma_{\chi p}^{S.I.}\propto
  \frac{1}{M_H^4}$. For large values of $\tan\beta$, the enhanced coupling
of the Higgs to $d$-type quarks tends to further increase the
cross section. Moreover because of the low $\mu$ values that these
models admit, the LSP coupling to the Higgses is maximised.

\begin{figure*}[hbtp]
\begin{center}
\mbox{\includegraphics[width=8cm,height=8cm]{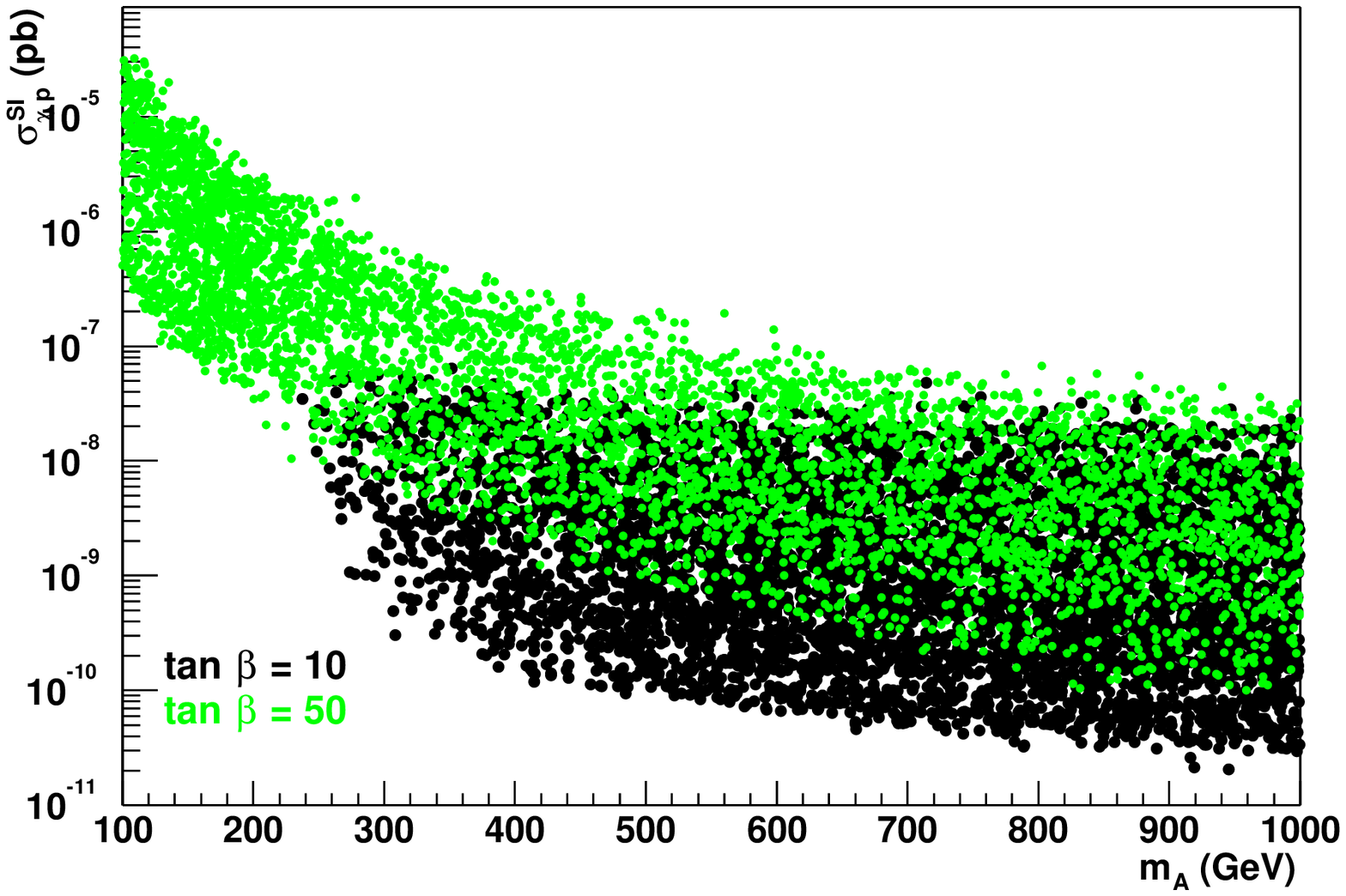}
\includegraphics[width=8cm,height=8cm]{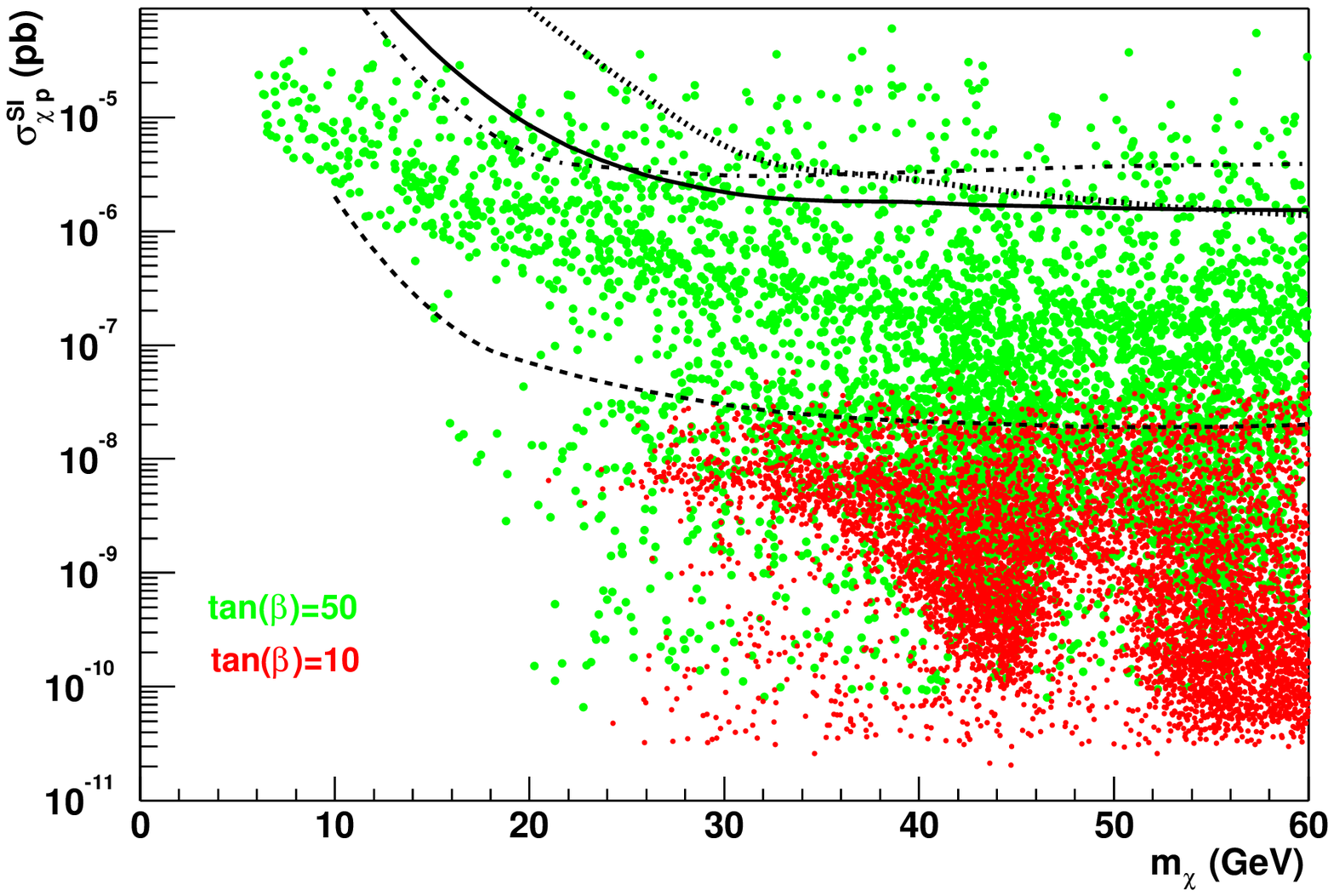}}
\vspace{-1cm} \caption{\label{si_ma}{\em  The spin independent
neutralino-proton cross section  as a function of a) $\ma$ and b)
$\mneuto$ for $\tan\beta=10$ (dark grey) and $\tan\beta=50$
(green/light grey). Scan over the parameters specified in Eq.
\ref{scan}. All constraints were imposed. The reach of the direct
detection experiments {\tt ZeplinI}(full), {\tt Edelweiss}(dot)
{\tt CDMS}(dot-dash) as well as for {\tt ZeplinII}(dash) is
displayed. }}
\end{center}
\end{figure*}

We have calculated the scalar cross section for neutralino scattering
off nucleons. We have included the contribution of all six flavors of
 quarks. Our results agree with the ones given in Ref.~\cite{direct_ellis,direct_vergados}.
We have used the values for the coefficients of the quark mass operator in the proton,
$f_{Tq}^{(p)}$ as given in Ref.~\cite{direct_ellis}.
  We show the predictions for the spin-independent cross section
for neutralino-proton scattering as a function of $\ma$,
Fig.~\ref{si_ma}a, after scanning over the parameter space as
specified in Eq.~\ref{scan}. One witnesses the large enhancement at low
values of $\ma$.   The various constraints from LEP, cosmology as
well as from precision measurements have been folded in the global
scan over the parameter space, hence the much smaller predictions
for the cross section at intermediate values of $\tan\beta$.
There, as was discussed above, the $\bsgamma$ constraint is not
compatible with light pseudoscalars. The cross section for
$\tan\beta=10$ therefore never exceeds a few $10^{-8}$pb.

For $\tan\beta=50$,  one reaches cross sections that are  in the
region of detectability of {\tt CDMS/Edelweiss/Zeplin} although
the largest cross sections are often found in regions where
neutralinos are very light, precisely where detectors have shown
less sensitivity, Fig.~\ref{si_ma}b. Nevertheless a few models,
albeit not the ones with the LSP near the lowest limit $\approx
6$GeV, are already ruled out by {\tt CDMS} and {\tt
Zeplin}\cite{zeplin}. Furthermore, in the majority of models with
$\ma<200 $GeV, the spin-independent cross section exceeds
$10^{-8}$, detectable by the next generation of direct dark matter
detectors such as {\tt EdelweissII} or {\tt ZeplinII}.{\footnote{
Note that there are uncertainties in the  exclusion curves written
in terms of the LSP scattering cross section on proton, for a
discussion of the effect of
 halo modeling and LSP  velocity distribution see\cite{halo-modeling-wimp}.} We emphasize
that to cover all models it is crucial to have a good sensitivity
in the region $\mneuto\approx 6-10$~GeV and that future detectors should extend their
capabilities and searches in this mass range.

\begin{figure*}[hbtp]
\begin{center}
\includegraphics[width=14cm,height=10cm]{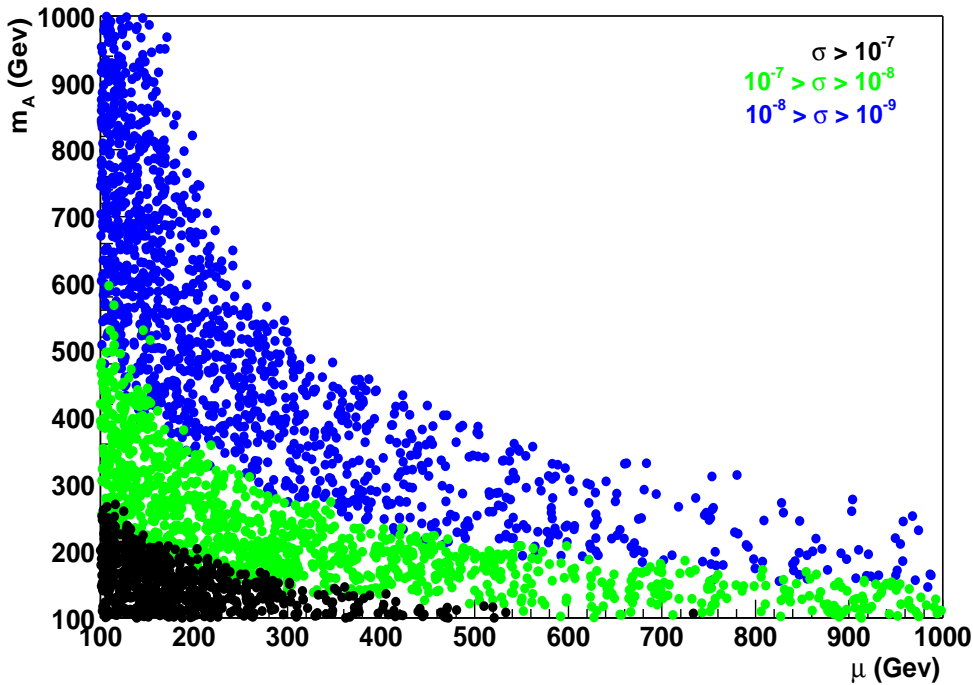}
\vspace{-.1cm} \caption{\label{t50_mamu}{\em  Regions of the
$\ma-\mu$ plane where  the spin independent neutralino-proton
cross section exceeds $10^{-7}$pb (black),  $10^{-8}$pb
(green/light grey), $10^{-9}$pb (blue/medium grey). Here,
$\tan\beta=50$ and we  scanned over the parameters specified in
Eq. \ref{scan}. All constraints were imposed. }}
\end{center}
\end{figure*}

Although the spin independent cross sections are enhanced at large
$\tan\beta$, cross sections above $10^{-7}$pb are found for all
values of $\tan\beta$ where light pseudoscalars are allowed
($\tan\beta>25$). Isocurves of the scalar cross section are
displayed in the $\ma-\mu$ plane in Fig.~\ref{t50_mamu} emphasizing
the importance of a sizable amount of gaugino-Higgsino mixing to
obtain large spin-independent cross sections.
Here only the upper bound on the relic density has been
considered. Many of the models, near the $Z$ or Higgs resonance
actually have a value for the relic density that is much too low.
Usually one rescales the cross section by a factor of the relic
density to some minimal value(say $0.05$) in order to take into
account the fact that the neutralino would not constitute the main
 cold dark matter component.
  Applying a rescaling factor of $\rho=\Omega
h^2/0.05$, would strongly affect the effective cross
section($\sigma_{eff}=\rho \sigma_{\chi p}^{S.I.})$ near the $Z$
or Higgs pole but not for the very light LSP where the relic
density is near the allowed upper bound.

\section{Prospects at $\epem$ colliders}

The aim of this section is to address the issue whether SUSY
models with non-universal gaugino masses that have a light LSP, say
$\mneuto<70$GeV, can always be probed at an $\epem$ linear
collider of centre of mass 500GeV. In the usual mSUGRA type model, imposing such a
light neutralino entails $\mchargo<150-200$ GeV  and therefore
independently of any other sparticles, chargino discovery is
guaranteed \cite{relic-sugra-old}.
 The non-universal models do not always ensure $\mchargo<250$~GeV,
 moreover some models do not  have light
sleptons as we have seen and yet are consistent with all data (including cosmology).

The main channels for producing supersymmetric particles at a
500GeV linear collider are the slepton pair production and
chargino pair production with $\epem\ra \neuto\tilde{\chi}^0_j$
being important processes as well. Of course the lightest Higgs
boson can always be produced. In non universal models, the
difficult cases for the LC will be the ones where sleptons are
heavy with the mass of the LSP near $M_Z/2$\footnote{These models
should not be thought of as fined tuned, from the relic density
point of view, more than the mSUGRA model is. The latter for
example does require, for example, near degeneracy between the
$\stau_1$ and $\neuto$ masses to pass the {\tt WMAP} constraint.}
. In this case large values of $\mu$ are still allowed.
This entails  charginos  (and $\neutt$) too heavy to be directly pair produced.
The only processes accessible would then be $\neuto\tilde{\chi}^0_j$

To ascertain which (s)particles can be reached at a 500 GeV,
  we perform scans over the parameters
$M_2,\mu,\m0$ for fixed  values of $\tan\beta$ and $\r12$, the non
universality parameter. We  compute the unpolarized cross sections
for $\epem\ra \ser\ser,{\tilde{\mu}_R}{\tilde{\mu}_R},{\tilde
\tau}_1{\tilde \tau}_1,\chargop\chargom,
\neuto\neutt,\neuto\neutth$ as well as for $\neuto\neuto\gamma$.
We have not simulated $\tilde{e}_L \ser$ and $\tilde{e}_L
\tilde{e}_L$ production. In our scenario these would be accessible
only when  $\ser \ser$ is accessible. The cross sections  are computed
with CalcHEP\cite{calchep,comphep_susymodel}, a program for the
automatic calculation of Feynman diagrams in the standard model or
the MSSM. Because the LSP is very light, clean signatures for all
sparticles that decay into $\neuto$  are guaranteed. In fact,
there is in general sufficient phase space for the decay into LSP
to constitute  the main decay mode for the selectron, $\ser\ra
e^-\neuto$,
 the chargino $\chargop\ra W^+ \neuto$ as well as heavy neutralinos,
$\tilde{\chi}_i\ra\neuto Z^{(*)}$. For charginos and neutralinos,
 decays into sleptons are also sometimes kinematically accesssible.
We will not perform a detailed analysis of signal and background
for all these processes,  in the $e^+e^-$ clean environment it
should not be a problem to see a signal for $\sigma> 1$fb.

We also consider $\epem\ra\neuto\neuto\gamma$, the process
that contribute to  $\siginv$. Although this process has to fight
a large background from the standard $\epem\ra\nu\bar\nu\gamma$,
with 95\% $e^-$ polarization and after a cut on the small angle
photons, it was shown that it was possible to reduce the
background to $<150 $fb.\cite{baer-lsplspgamma} From this we
estimate that with a luminosity of $500fb^{-1}$ , the polarized
cross section for the signal should exceed $\sigma=1.6$~fb in
order to see a $3\sigma$ deviation. For simplicity we compute
only the unpolarised cross section and impose  conservatively the
same value as the pair production processes, $\sigma_{\rm
unpol}>1$fb.

\begin{figure*}[hbtp]
\begin{center}
\includegraphics[width=15cm,height=14cm]{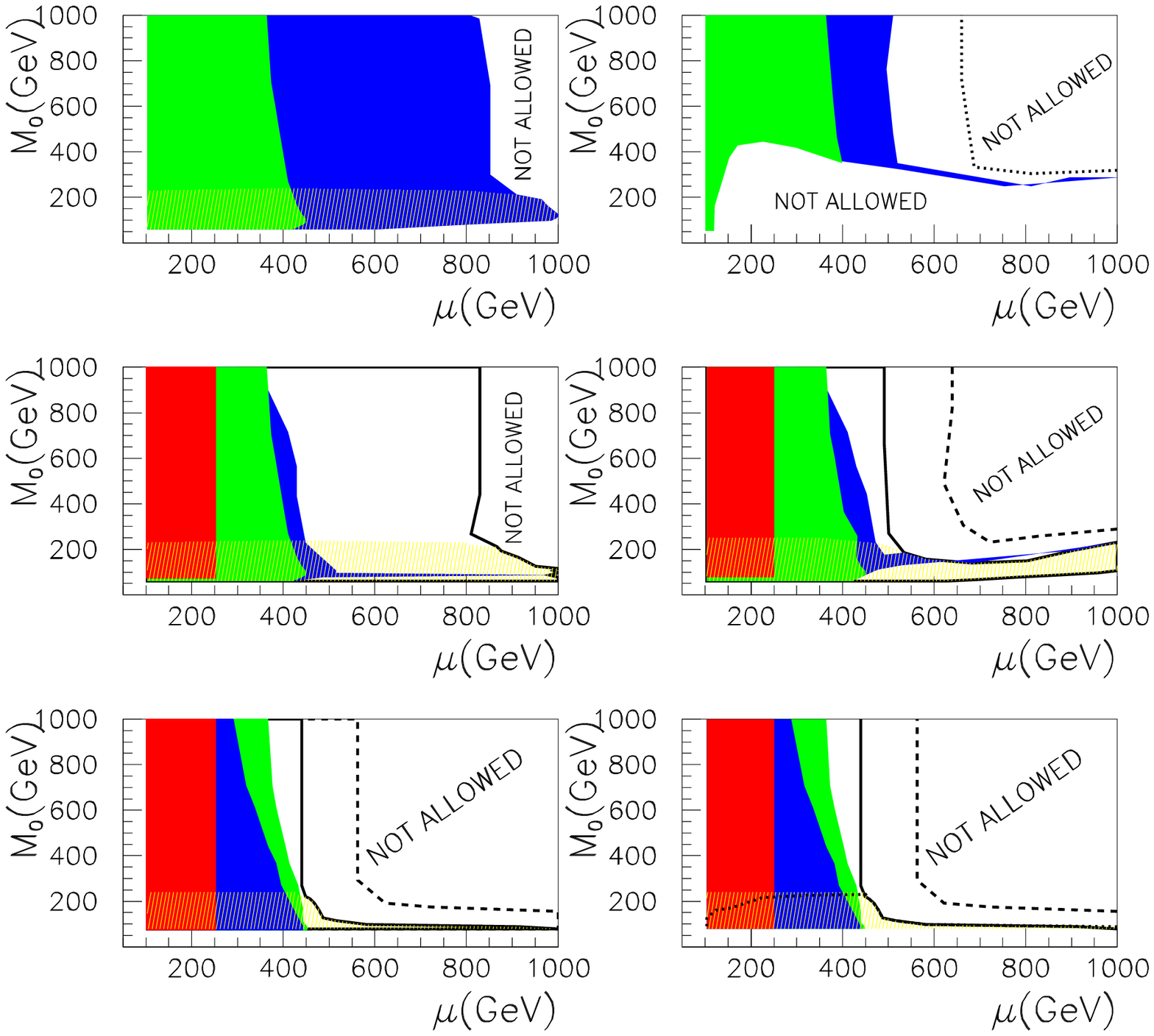}
\vspace{-1cm} \caption{\label{t10ee}{\em  Region of $\m0$-$\mu$
plane where the cross section for sparticles production in $\epem$
exceeds 1fb, for $\tan\beta=10$(left) and $\tan\beta=50$ (right)
and from top to bottom $\r12=0.5,0.1,0.03$.  $\r12=0.5$ is the
usual GUT value. The processes are $\ser\ser$ (yellow hatch)
$\chargop\chargom$ (red/medium grey) $\neuto\neutt$ (blue/dark
grey) $\neuto\neutth$ (green/light grey). Note that there is a
large overlap between the region covered by $\ser\ser$ and the
ones (not featured) covered by $\smur\smur$ or $\stau_1\stau_1$.
The region to the right of the  full (dashed) line do not satisfy
the constraint $\Omega h^2<0.128(0.3)$ coupled with {\tt LEP} and
precision measurements. In the region (white) to the left of the
{\tt WMAP} line no supersymmetric particle can be produced. Such
regions are present only when $\r12=0.03$ and $\r12=0.1$. In the
top figures, the whole parameter space is also covered by
$\neuto\neutt$.
 In the bottom right figure, the dotted line delimits
 the region where $\sigma(\epem\ra\neuto\neuto\gamma)>1$fb.}
}
\end{center}
\end{figure*}

Here we concentrate on the $\mu>0$ region. We find that in large
regions of the parameter space the cross section for at least one
process exceeds 1fb. We do, however, uncover regions compatible
with existing limits
 where no sparticle is
produced.
 To look more precisely at the potential of a 500GeV
LC, we discuss again two typical values of $\tan\beta$.

\noi
$\bullet \tan\beta=10$

First consider the case $\tan\beta=10$. For  unified models
($r_{12}\approx .5$) and with $\mneuto<70$~GeV, the
whole parameter space is covered basically by the  processes
$\epem\ra\chargop\chargom$.  While chargino pair production is the preferred channel
especially for the large $\m0$ region, in this
scenario, $\neutt\neuto$ also exceeds 1fb in the full parameter
space, Fig.~\ref{t10ee}.
The process   $\neutth\neuto$ has a significant cross section only for $\mu\leq 400$GeV.
When $\m0<200$GeV, the slepton pair
production is also always accessible.
 As
one decreases the value of the non universality parameter,
$r_{12}=0.1$, one starts uncovering permitted regions where none
of the charginos/neutralinos production processes have a large
enough cross section while the sleptons are too heavy for pair
production. This occurs in the large $\m0$ region and for very
large $\mu$ where the coupling to the $Z$ is not sufficient. As
the chargino mass can be much higher than the LSP, the chargino
process is accessible only in the small $\mu$ region while the
$\neuto\neutt$ and $\neuto\neutth$ both extend the reach in
parameter space. The process $\neuto\neutt$ is also observable in
a narrow region at small $\m0$ even for large values of $\mu$,
these were the points where $\mneuto\approx M_Z/2$.

When $\r12$ decreases to $\r12=0.03$, one observes the same
features. For $\m0<200$GeV, observable cross sections for the
slepton production processes $\ser\ser,\smur\smur$ and
$\stau_1\stau_1$ are expected as for the universal case.
 For larger values of $\m0$, it is $\sigma_{\neuto\neutth}$ that
eventually exceeds $\sigma_{\neuto\neutt}$ in the region
$M_1<<\mu<M_2$. Then the LSP is  bino-like and both $\neutt$ and
$\neutth$ are Higgsino-like and nearly degenerate. However the
$Z\neuto\neutth$  features the largest coupling.
With the new data from {\tt WMAP}, the large $\m0/\mu$
region is more severely constrained,
 thus reducing the allowed
parameter space where no supersymmetric particles can be produced.

Note that one can also get indirect evidence of light neutralinos
by measuring the invisible branching fraction of the Higgs. A
large branching fraction into invisible is expected  in models
with non-universal gaugino masses in the region of intermediate
$\tan\beta$ and when $\mu$ is small \cite{nous_hinvisible_lhc}.
For example, for $\tan\beta=10, \r12=0.1, \mu<300$~GeV the
branching fraction $h\ra\neuto\neuto$ can reach $60\%$. This
region of parameter space corresponds to the one where one expects
many other signals of supersymmetry at the LC500,
$\chargop\chargom,\neuto\tilde{\chi}_j$ and sometimes $\ser\ser$,
Fig.~\ref{t10ee}. Correlating the information from the Higgs and
the supersymmetric sector will then help  to establish the nature
of the invisible decay mode of the Higgs.

\noi
$\bullet \tan\beta=50$

In the large $\tan\beta$ scenario, the slepton pair production
process is not so useful, at least in the unified model,
 as most of the low $\m0$ region is ruled out.
The selectron pair production process is only accessible in a
corner of parameter space when $\mu,\m0\approx 100$GeV. The stau pair
production is also accessible when  $\mu$ is large enough to
induce significant mixing in the $\tilde\tau$ sector. However,
with the {\tt WMAP} constraint there remains only a very narrow region
at large $\mu$.
Both chargino pairs and $\neuto\neutt$
cover all allowed parameter space while $\neuto\neutth$ is also important when
$\mu<400$~GeV.
 As
$\r12$ decreases, the low $\m0$ region is allowed and observable
cross sections for $\ser\ser$ as well as $\stau_1\stau_1$ are
predicted. Again  there is a portion of the  large $\m0$-large
$\mu$ region of the parameter space where supersymmetric particles
are too heavy to be produced. Note that imposing the new {\tt WMAP}
constraint has considerably shrunk this region.
Finally,  the cross section
for the process $\neuto\neuto\gamma$ can exceed $1$fb, although
this occurs only in the low $\m0$ region when slepton pair production is also
accessible.

Increasing the energy of the collider would obviously increase the
parameter space where sparticles could be pair produced.
Basically, the slepton pair production is large for $\m0<400$GeV
and chargino pair production for $\mu<450$GeV. Thus, one nearly
recovers full coverage  with only these two processes,
 in the case of nonuniversal models with  light neutralinos
 at a 800GeV collider. At first sight, the more problematic models are those
 with   $\r12\approx 0.1$ where much larger values of $\mu$ are allowed
 at intermediate $\tan\beta$.  However,  the process $\epem\ra\neuto\neutt$ has
 a reasonable cross section for most of the large $\m0-\mu$ region.
 For example when $\tan\beta=10$,  $\neuto\neutt$ is measurable everywhere except
 in a small  region where  $\m0>900$GeV.
  Then both the  $t$-channel sfermion exchange and the
$s$-channel $Z$ exchange  contributions are small and no signal
for supersymmetric particles  can be expected. With {\tt WMAP},
this difficult region is already ruled out when $\r12<0.1$ (see
Fig.~\ref{t10ee} for the allowed region). One also  avoids the
difficult region when $\tan\beta=50$ (Fig.~\ref{t10ee}). There, for
any value of the non universality parameter the allowed region
corresponds roughly to the one where chargino pair production is
accessible.

Finally we  mention that such light supersymmetric particles would
probably first  be discovered at the hadron colliders. For the
{\tt Tevatron} , although the chargino mass considered are often
within the  range where discovery via the trilepton signal is
expected ($\mchargo$ up to 250 GeV) one needs to take a  closer
look at cross sections for signal and backgrounds in nonunified
models to ascertain the viability of the signal
\cite{tri-leptons-nonuni}. In the models with heavy squarks and
Higgs scalars, no other opportunity for SUSY discovery exists. The
LHC on the other hand would have plenty of opportunities to
discover SUSY in either the squark, Higgs or gaugino sector. We
should also note that the LC through precision measurements on the
$h$ properties could also probe into SUSY especially by using
information from the LHC.

\section{Conclusion}

We have reexamined the lowest bound on the neutralino LSP in the
minimal supersymmetric model. We have worked within the context of
minimal flavour violation and $R$ parity conserving
supersymmetric models. We have reduced the parameter
 space to only a few important parameters,
 those of the gaugino and slepton sector, the pseudoscalar mass and the
 trilinear coupling of the squarks.
In particular we have relaxed the universality relation between
the gaugino masses thus removing the most important constraint on
the LSP mass arising from LEP. We find that the upper limit on the
relic density contributed by neutralinos and as inferred from the
new data by {\tt WMAP} basically sets the lower bound on
neutralinos in models where the gaugino masses are not unified at
the high scale but satisfy $M_1<<M_2$. In the limit of a heavy
pseudoscalar mass, we found a lower bound on the neutralino mass
of $18$GeV in models with  $\tan\beta=50$. For intermediate
values of $\tan\beta$, the annihilation of neutralinos into light
fermions is not as efficient and one can only set a bound of
$29$GeV for $\tan\beta=10$. It is however in models  with light
pseudoscalar masses ($\ma< 200$GeV) and large values of
$\tan\beta$ that one finds the lightest LSP $\mneuto\approx 6$GeV.
This is due to a new contribution to the annihilation
cross-section of neutralinos, scalar exchange mediating
$\neuto\neuto\ra \tau^+ \tau^-$, thus reducing the relic density.

The models with light pseudoscalars and neutralinos below $\approx
16$~GeV can be probed both in the next generation of direct
detection experiments as well as at the {\tt Tevatron}  or future
colliders. The cross section for the former are much enhanced for
a light Higgs with significant coupling to the neutralino. It is
however crucial that the detector be sensitive to light neutralino
masses. In addition, the pseudoscalar and/or scalar Higgses should
 be within reach of the {\tt Tevatron}  RunIIa.
 Charginos or $\bsmu$ might also be detected there.
 The LHC also has a potential for discovering charged Higgses in the large
$\tan\beta$ region
 using $gb\ra t H^\pm$ with $ H^+\ra t\bar{b},\tau^+\bar{\nu}$ or $qq\ra H^\pm\ra \tau\nu$ \cite{higgs_houches2003}.
 Studies of neutral Higgses searches in
the $\tau^+\tau^-$ channel, also show a good potential for discovery
even with a low luminosity \cite{higgs_houches2003}. Of course other sparticles, such as charginos, heavier
neutralinos or sfermions might be discovered as well.

Finally we have shown that
a linear collider with  centre of mass of $500$GeV has good prospects for producing
supersymmetric particles in models where neutralinos are below the
weak scale, although the parameter space cannot be completely
covered especially in models where $M_1<<M_2$. The gaugino
universality relation can be directly tested at such a collider
but this necessitates performing a combined fit to the neutralino
mass (measured in slepton pair production process), the chargino
mass as well as to the polarized cross sections for
$\epem_R\ra\ser\ser,\chargop\chargom$
\cite{JapanSusy,fawzi_india}. All these precision  measurements
can be realized only if $\m0(\mse1),\mu < 250$GeV.
Of course  models  where $\ma$ is near ${\cal O}(100)$ GeV should
lead to several signals at  the linear collider. Most noticeably
in the Higgs sector where one expects a large cross section
($10-100 fb$) in both  $\epem\ra ZH,hA$ channels as well as in the
charged Higgs pair production channel. In most cases, and most
importantly when the Higgs sector is harder to probe, the
charginos would be accessible as well at a collider with
$\sqrt{s}=500$GeV.

\noi {\bf Acknowledgments}

We thank A. Czarnecki for discussions on the theoretical calculation of \gmuon. We also thank A.~Semenov for the realization of the MSSM model file
implemented  in CalcHEP.
 This work was supported in part by the PICS-397 {\it Calculs automatiques en physique des particules}.

\end{document}